\input harvmac.tex

\vskip 1.5in
\Title{\vbox{\baselineskip12pt
\hbox to \hsize{\hfill}
\hbox to \hsize{\hfill }}}
{\vbox{
	\centerline{\hbox{Higher Spin Holography
		}}\vskip 5pt
        \centerline{\hbox{ and $AdS$ String Sigma-Model
		}} } }
\centerline{Dimitri Polyakov$^{}$\footnote{$^\dagger$}
{polyakov@sogang.ac.kr ;
twistorstring@gmail.com
}}
\medskip
\centerline{\it Center for Quantum Space-Time (CQUeST)$^{}$}
\centerline{\it Sogang University}
\centerline{\it Seoul 121-742, Korea}

\vskip .3in

\centerline {\bf Abstract}

We analyze 
cubic spin 3 interaction in AdS space
using the higher spin
extension of string-theoretic sigma-model 
constructed in our previous work, which low
energy limit is described by AdS vacuum solution.
We find that, in the leading order of  the cosmological constant,
the spin 3 correlator on the $AdS_4$ string theory side reproduces
the structure of 3-point function of composite operators,
quadratic in free fields, in the dual $d=3$  vector model.
 The cancellation of holography violating
terms in $d=3$ is related to the value of  the Liouville background
charge in $d=4$.

\Date{July 2012}

\vfill\eject
\lref\malda{J. Maldacena, 
Adv.Theor.Math.Phys. 2 (1998) 231-252}
\lref\wit{E. Witten, 
Adv.Theor.Math.Phys. 2 (1998) 253-291 }
\lref\klebpol{S. Gubser, I. Klebanov, A. M. Polyakov, Phys.Lett. B428 (1998) 105-114}
\lref\hvs{M. Vasiliev, arXiv:1203.5554}
\lref\maldaz{J. Maldacena, A. Zhiboedov, arXiv:1112.1016}
\lref\maldazh{J. Maldacena, A. Zhiboedov, arXiv:1204.3882}
\lref\antalf{. de Mello Koch, A. Jevicki, K. Jin, J. P. Rodrigues,
Phys.Rev. D83 (2011) 025006}
\lref\antals{	
R. de Mello Koch, A. Jevicki, K. Jin, J. P. Rodrigues, Q. Ye, arXiv:
1205.4117}
\lref\metsaevf{ R. Metsaev, arXiv: 1112.0976}
\lref\metsaevs{ R. Metsaev, arXiv:1205.3131}
\lref\tsul{P. Dempster, M. Tsulaia, arXiv:1203.5597}
\lref\selfp{S. Lee, D. Polyakov, Phys.Rev. D85 (2012) 106014}
\lref\selfpp{D. Polyakov, Phys.Rev. D84 (2011) 126004} 
\lref\fvf{E.S. Fradkin, M.A. Vasiliev, Nucl. Phys. B 291, 141 (1987)}
\lref\fvs{E.S. Fradkin, M.A. Vasiliev, Phys. Lett. B 189 (1987) 89}
\lref\mmswf{S.W. MacDowell, F. Mansouri, Phys. Rev.Lett. 38 (1977) 739}
\lref\mmsws{K. S. Stelle and P. C. West, Phys. Rev. D 21 (1980) 1466}
\lref\mmswt{C.Preitschopf and M.A.Vasiliev, hep-th/9805127}
\lref\vmaf{M. A. Vasiliev, Sov. J. Nucl. Phys. 32 (1980) 439,
Yad. Fiz. 32 (1980) 855}
\lref\vmas{V. E. Lopatin and M. A. Vasiliev, Mod. Phys. Lett. A 3 (1988) 257}
\lref\vmat{E.S. Fradkin and M.A. Vasiliev, Mod. Phys. Lett. A 3 (1988) 2983}
\lref\vmafth{M. A. Vasiliev, Nucl. Phys. B 616 (2001) 106 }
\lref\bonellio{G. Bonelli, JHEP 0411 (2004) 059}
\lref\hsaone{M. A. Vasiliev, Fortsch. Phys. 36 (1988) 33}
\lref\hsatwo{E.S. Fradkin and M.A. Vasiliev, Mod. Phys. Lett. A 3 (1988) 2983}
\lref\hsathree{S. E. Konstein and M. A. Vasiliev, Nucl. Phys. B 331 (1990) 475}
\lref\hsafour{M.P. Blencowe, Class. Quantum Grav. 6, 443 (1989)}
\lref\hsafive{E. Bergshoeff, M. Blencowe and K. Stelle, 
Comm. Math. Phys. 128 (1990) 213}
\lref\hsasix{E. Sezgin and P. Sundell, Nucl. Phys. B 634 (2002) 120 }
\lref\hsaseven{M. A. Vasiliev, Phys. Rev. D 66 (2002) 066006 }
\lref\soojongf{M. Henneaux, S.-J. Rey, JHEP 1012:007,2010}
\lref\henneaux{J. D. Brown and M. Henneaux, Commun. Math. Phys. 104, 207 (1986)}
\lref\sagnottia{A. Sagnotti, E. Sezgin, P. Sundell, hep-th/0501156}
\lref\fronsdal{C. Fronsdal, Phys. Rev. D18 (1978) 3624}
\lref\coleman{ S. Coleman, J. Mandula, Phys. Rev. 159 (1967) 1251}
\lref\haag{R. Haag, J. Lopuszanski, M. Sohnius, Nucl. Phys B88 (1975)
257}
\lref\weinberg{ S. Weinberg, Phys. Rev. 133(1964) B1049}
\lref\tseytbuch{E. Buchbinder, A. Tseytlin, JHEP 1008:057,2010}
\lref\fradkin{E. Fradkin, M. Vasiliev, Phys. Lett. B189 (1987) 89}
\lref\skvortsov{E. Skvortsov, M. Vasiliev, Nucl.Phys.B756:117-147 (2006)}
\lref\mva{M. Vasiliev, Phys. Lett. B243 (1990) 378}
\lref\mvb{M. Vasiliev, Int. J. Mod. Phys. D5
(1996) 763}
\lref\mvc{M. Vasiliev, Phys. Lett. B567 (2003) 139}
\lref\brink{A. Bengtsson, I. Bengtsson, L. Brink, Nucl. Phys. B227
 (1983) 31}
\lref\deser{S. Deser, Z. Yang, Class. Quant. Grav 7 (1990) 1491}
\lref\bengt{ A. Bengtsson, I. Bengtsson, N. Linden,
Class. Quant. Grav. 4 (1987) 1333}
\lref\boulanger{ X. Bekaert, N. Boulanger, S. Cnockaert,
J. Math. Phys 46 (2005) 012303}
\lref\bbd{F. Berends, G. Burgers, H. Van Dam ,Nucl.Phys. B260 (1985) 295}
\lref\metsaev{ R. Metsaev, arXiv:0712.3526}
\lref\siegel{ W. Siegel, B. Zwiebach, Nucl. Phys. B282 (1987) 125}
\lref\siegelb{W. Siegel, Nucl. Phys. B 263 (1986) 93}
\lref\nicolai{ A. Neveu, H. Nicolai, P. West, Nucl. Phys. B264 (1986) 573}
\lref\damour{T. Damour, S. Deser, Ann. Poincare Phys. Theor. 47 (1987) 277}
\lref\sagnottib{D. Francia, A. Sagnotti, Phys. Lett. B53 (2002) 303}
\lref\sagnottic{D. Francia, A. Sagnotti, Class. Quant. Grav.
 20 (2003) S473}
\lref\sagnottid{D. Francia, J. Mourad, A. Sagnotti, Nucl. Phys. B773
(2007) 203}
\lref\labastidaa{ J. Labastida, Nucl. Phys. B322 (1989)}
\lref\labastidab{ J. Labastida, Phys. Rev. Lett. 58 (1987) 632}
\lref\mvd{L. Brink, R.Metsaev, M. Vasiliev, Nucl. Phys. B 586 (2000) 183}
\lref\klebanov{ I. Klebanov, A. M. Polyakov,
Phys.Lett.B550 (2002) 213-219}
\lref\mve{
X. Bekaert, S. Cnockaert, C. Iazeolla,
M.A. Vasiliev,  IHES-P-04-47, ULB-TH-04-26, ROM2F-04-29, 
FIAN-TD-17-04, Sep 2005 86pp.}
\lref\sagnottie{A. Campoleoni, D. Francia, J. Mourad, A.
 Sagnotti, Nucl. Phys. B815 (2009) 289-367}
\lref\sagnottif{
A. Campoleoni, D. Francia, J. Mourad, A.
 Sagnotti, arXiv:0904.4447}
\lref\selfa{D. Polyakov, Int.J.Mod.Phys.A20:4001-4020,2005}
\lref\selfb{ D. Polyakov, arXiv:0905.4858}
\lref\selfc{D. Polyakov, arXiv:0906.3663, Int.J.Mod.Phys.A24:6177-6195 (2009)}
\lref\selfd{D. Polyakov, Phys.Rev.D65:084041 (2002)}
\lref\spinself{D. Polyakov, Phys.Rev.D82:066005,2010}
\lref\spinselff{D. Polyakov,Phys.Rev.D83:046005,2011}
\lref\mirian{A. Fotopoulos, M. Tsulaia, Phys.Rev.D76:025014,2007}
\lref\extraa{I. Buchbinder, V. Krykhtin,  arXiv:0707.2181}
\lref\extrac{X. Bekaert, I. Buchbinder, A. Pashnev, M. Tsulaia,
Class.Quant.Grav. 21 (2004) S1457-1464}
\lref\extraf{I. Buchbinder, A. Pashnev, M. Tsulaia, 
Phys.Lett.B523:338-346,2001}
\lref\extrag{I. Buchbinder, E. Fradkin, S. Lyakhovich, V. Pershin,
Phys.Lett. B304 (1993) 239-248}
\lref\bonellia{G. Bonelli, Nucl.Phys.B {669} (2003) 159}
\lref\bonellib{G. Bonelli, JHEP 0311 (2003) 028}
\lref\hsself{D.Polyakov, arXiv:1005.5512}
\lref\sundborg{ B. Sundborg, ucl.Phys.Proc.Suppl. 102 (2001)}
\lref\sezgin{E. Sezgin and P. Sundell,
Nucl.Phys.B644:303- 370,2002}
\lref\giombif{S. Giombi, Xi Yin, arXiv:0912.5105}
\lref\giombis{S. Giombi, Xi Yin, arXiv:1004.3736}
\lref\bekaert{X. Bekaert, N. Boulanger, P. Sundell, arXiv:1007.0435}
\lref\taronna{A. Sagnotti, M. Taronna, arXiv:1006.5242, 
Nucl.Phys.B842:299-361,2011}
\lref\fotopoulos{A. Fotopoulos, M. Tsulaia, arXiv:1007.0747}
\lref\fotopouloss{A. Fotopoulos, M. Tsulaia, arXiv:1009.0727}
\lref\taronnao{M. Taronna, arXiv:1005.3061}
\lref\taronnas{A. Sagnotti, M. Taronna, arXiv:1006.5242 ,
Nucl.Phys.B842:299-361,2011}
\lref\campo{A.Campoleoni,S. Fredenhagen,S. Pfenninger, S. Theisen,
arXiv:1008.4744, JHEP 1011 (2010) 007}
\lref\gaber{M. Gaberdiel, T. Hartman, arXiv:1101.2910, JHEP 1105 (2011) 031}
\lref\per{	
N. Boulanger,S. Leclercq, P. Sundell, JHEP 0808(2008) 056 }
\lref\mav{V. E. Lopatin and M. A. Vasiliev, Mod. Phys. Lett. A 3 (1988) 257}
\lref\zinov{Yu. Zinoviev, Nucl. Phys. B 808 (2009)}
\lref\sv{E.D. Skvortsov, M.A. Vasiliev,
Nucl. Phys.B 756 (2006)117}
\lref\mvasiliev{D.S. Ponomarev, M.A. Vasiliev, Nucl.Phys.B839:466-498,2010}
\lref\zhenya{E.D. Skvortsov, Yu.M. Zinoviev, arXiv:1007.4944}
\lref\perf{N. Boulanger, C. Iazeolla, P. Sundell, JHEP 0907 (2009) 013 }
\lref\pers{N. Boulanger, C. Iazeolla, P. Sundell, JHEP 0907 (2009) 014 }
\lref\selft{D. Polyakov,Phys.Rev.D82:066005,2010}
\lref\selftt{D. Polyakov, Int.J.Mod.Phys.A25:4623-4640,2010}
\lref\tseytlin{I. Klebanov, A Tseytlin, Nucl.Phys.B546:155-181,1999}
\lref\rubenf{R. Manvelyan, K. Mkrtchyan, W. Ruehl, Nucl.Phys.B836:204-221,2010}
\lref\robert{
R. De Mello Koch, A. Jevicki, K. Jin, J. A. P. Rodrigues, arXiv:1008.0633}
\lref\bekae{X. Bekaert, S. Cnockaert, C. Iazeolla, M. A. Vasiliev,
hep-th/0503128}
\lref\vcubic{M. Vasiliev, arXiv:1108.5921}
\lref\sagnottinew{A. Sagnotti, arXiv:1112.4285}
\lref\yin{C.-M. Chang, X. Yin, arXiv:1106.2580 }
\lref\boulskv{ N. Boulanger, E. Skvortsov, arXiv:1107.5028,
JHEP 1109 (2011) 063}
\lref\boulskvz{N. Boulanger, E. Skvortsov,Yu. Zinoviev,
arXiv:1107.1872 , J.Phys.A A44 (2011) 415403
}
\lref\selfsigma{D. Polyakov, Phys.Rev. D84 (2011) 126004}
\lref\wittwist{E. Witten, Commun.Math.Phys. 252 (2004) 189-258}
\lref\soojongs{M. Henneaux, G. L. Gomez, J. Park, S.-J. Rey, 
JHEP 1206 (2012) 037}
\lref\joung{E. Joung, L. Lopez, M. Taronna, JHEP 1207 (2012) 041}
\lref\sor{D. Sorokin, AIP
Conf. Proc. 767 (2005) 172 [hep-th/0405069].
}
\lref\bek{ X. Bekaert, N. Boulanger and Per A. Sundell, Rev. Mod. Phys. 84 (2012) 987
}
\lref\sez{E. Sezgin and P. Sundell, JHEP 0507 (2005) 044 [hep-
th/0305040]
}
\lref\mtar{M. Taronna, JHEP 1204 (2012) 029}

\centerline{\bf  1. Introduction}

It is common to think of $AdS/CFT$ holography
as of duality between semiclassical limit of supergravity
with negative cosmological constant and  conformal field theory
(CFT) living on the boundary of its vacuum solution (AdS space).
This, however, is the low energy approximation;
in the stronger sense the $AdS/CFT$ conjecture
means that the correlation functions  of physical vertex 
operators computed in closed string theory in anti-de Sitter
background must reproduce the correlators of the corresponding
conformally invariant observables on the CFT side 
~{\malda, \klebpol, \wit}.
Regardless of space-time dimension, higher spin fields in
AdS space (with various symmetries) inevitably
have to play critical role in AdS/CFT holography since the
overwhelming number of operators  on the CFT (gauge theory side),
for example, those of the type 
\eqn\lowen{\sim{Tr(\phi_I\nabla_{m_1}...\nabla_{m_s}\phi^I)}}
simply have no choice but to match the higher  spin objects
propagating in AdS space-time (and possibly polarized along 
the direction of the boundary). 
In particular, it has been conjectured ~{\sezgin, \klebanov}
 that , in case
of $AdS_4/CFT_3$, the symmetric fields of spin $s$ in $AdS_4$
described by Vasiliev's unfolding formalism 
(e.g. see ~{\fvf, \fvs, \vmaf, \vmas, \vmat, \hsaone, \hsatwo} are dual
to the symmetrized objects of the type (1) at the
conformal points of the $O(N)$ vector
model in $d=3$ for even values of $s$ and 
the $U(N)$ model
for odd spins. This conjecture has been checked explicitly
in important papers  ~{\giombif, \giombis, \yin}
 and later analyzed in a number of insightful works, e.g. 
 ~{\maldaz, \maldazh, \hvs, \antalf, \antals}
whose results suggest the importance of the free field
theory limit in the $O(N)/HS$ duality, despite the fact
that Maldacena-Zhiboedov theorem can be circumvented
under certain assumptions ~{\hvs}.
Apart from the low-energy limit,
the dynamics of these higher spin fields is described by physical
vertex operators  in open or closed string theories in anti-de Sitter
space and their worldsheet correlation functions.
In particular, the $AdS/CFT$ duality conjecture
strongly suggests the existence of infinite tower of $massless$
higher spin states in the string spectrum in $AdS$ space-time.
In practice, however, little is known about AdS string theory
dynamics beyond semiclassical limit, since straightforward
quantization of string theory in AdS space-time is not known 
(e.g. see ~{\tseytbuch}). 
Another important point is that, in the standard description,
the string excitations correspond to the space-time fields
in the metric ~{\fronsdal, \sor, \bek, \sagnottia,
\sagnottib, \sagnottic, \sagnottid, \taronnas,
\labastidaa, \labastidab} 
rather than unfolded formulation,
while it is the unfolded formalism which is the
most natural and efficient frame-work  to approach
the problem of the higher spin extension of the $AdS/CFT$ duality
~{ \sezgin, \klebanov, \giombif, \giombis, \yin, \hvs, \sez}.
In one of the recent works ~{\selfpp} we constructed the 
string-theoretic sigma model based on hidden space-time symmetry
generators in RNS formalism, realizing $AdS_d$ isometry group.
The model is initially  defined in the
flat background, however, when perturbed by the vertex operators
based on the hidden $AdS$ isometry generators, it flows to the new 
fixed $2d$ conformal  point, corresponding to $AdS$ geometry in 
space-time. This can be shown by analyzing the conformal beta-function
of the sigma-model, resulting in the low-energy effective equations
of motion, describing (in the leading order)  the $AdS_d$
 vacuum solutions of gravity with negative cosmological constant in the
Mac Dowell-Mansouri-Stelle-West description ~{\mmswf, \mmsws, \mmswt} 
 Remarkably,
the closed string vertex operators, constructed in ~{\selfpp}
describe the gravitational excitations around the AdS vacuum in the
frame, rather than metric formalism - i.e. in terms of the
vielbein and spin connection gauge fields.
In this paper we extend our analysis 
of this sigma-model to include the excitations corresponding
to massless higher spin fields in the frame-like Vasiliev's approach.
The string-theoretic vertex operators for the frame-like higher spin
fields have been constructed in our earlier work ~{\selfp} where we 
performed their BRST analysis and analyzed their correlators
in flat space, showing them to lead to Berends - Burgers - Van Dam
type  {\bbd} cubic spin 3 interactions 
~{\metsaev, \metsaevf, \metsaevs, \bek, \boulanger,
\joung, \mirian, \extraa, \zhenya, \perf, \pers, 
\mtar, \rubenf}
in flat space. In the current paper
we extend this analysis using the sigma-model approach ~{\selfpp}
  in order
to study the AdS deformations  of these cubic interactions
and their relevance to higher spin holography problem
~{\hvs, \giombif, \giombis, \yin, \maldaz, \maldazh, \antalf, \antals, \soojongf, \soojongs,
\campo, \gaber}
, in the context
of ~{\klebanov}. 
We find that, in the leading nontrivial order in $\rho^{-1}$,
the cubic spin $3$ interaction reproduces 
the correlators of the operators of type (1) in the free field
limit of the $U(N)$ model in $d=3$.
 In particular, in this limit the cubic spin 3 interaction is
dominated by the $9$-derivative terms, while the lower derivative
terms  (posing a potential threat to the holography) are absent 
as their cancellation is ensured by the ghost number
selection rules  for the vertex operators and by
the value of the Liouville
background charge ($q={\sqrt{5\over2}}$) in $d=4$. The terms
with the lower number of derivatives, however, are generally present
in the sigma-model for $d\neq{4}$. In addition, in the  $d=4$ case
these terms
may still appear in the higher order corrections in $\alpha^{\prime}$
(corresponding to ${1\over{N}}$ corrections in the dual theory).
Next, it is the momentum behaviour
and the  pole structure of the  string-theoretic
spin $3$  amplitude in the sigma-model
that corresponds to the field-theoretic pole structure
of the 3-point amplitude of the operators (1)
of the dual theory in the momentum space.
In the following sections, we shall 
use the AdS string sigma-model to perform the explicit computation
of the $3$-point correlators using the vertex operators for spin
3 fields in the frame-like formalism (in the leading nontrivial
orders in $\Lambda$ and $\alpha^\prime$) and discuss physical
implications of our results.

\centerline{\bf 2. Sigma-Model for AdS Strings and Vertex Operators}

\centerline{\bf for Frame-like Higher Spin Fields: a Brief Review}

The sigma-model for AdS strings constructed in ~{\selfpp}
 is based on
hidden space-time symmetry generators in RNS superstring theory.
Namely, consider the RNS superstring theory in flat space with the
 action given by:
\eqn\grav{\eqalign{S_{RNS}=S_{matter}+S_{bc}+S_{\beta\gamma}+S_{Liouville}\cr
S_{matter}=-{1\over{4\pi}}\int{d^2z}(\partial{X_m}\bar\partial{X^m}
+\psi_m\bar\partial\psi^m+{\bar\psi}_m\partial{\bar\psi}^m)\cr
S_{bc}={1\over{2\pi}}\int{d^2z}(b\bar\partial{c}+{\bar{b}}\partial
{\bar{c}})\cr
S_{\beta\gamma}={1\over{2\pi}}\int{d^2z}(\beta\bar\partial\gamma
+\bar\beta\partial{\bar\gamma})
\cr
S_{Liouville}=-{1\over{4\pi}}\int{d^2z}(\partial\varphi\bar\partial\varphi
+\bar\partial\lambda\lambda+\partial\bar\lambda\bar\lambda
+\mu_0{e^{B\varphi}}(\lambda\bar\lambda+F))
}}
where $\varphi,\lambda, F$ are components of super Liouville field
and the Liouville background charge is
\eqn\lowen{
q=B+B^{-1}={\sqrt{{{9-d}\over2}}}}

The ghost fields $b,c,\beta,\gamma$ are bosonized according to

\eqn\grav{\eqalign{b=e^{-\sigma},c=e^{\sigma}\cr
\gamma=e^{\phi-\chi}\equiv{e^\phi}\eta\cr
\beta=e^{\chi-\phi}\partial\chi\equiv\partial\xi{e^{-\phi}}}}

and the BRST charge is
\eqn\grav{\eqalign{Q=Q_1+Q_2+Q_3\cr
Q_1=\oint{{dz}\over{2i\pi}}(cT-bc\partial{c}) \cr
Q_2=-{1\over2}\oint{{dz}\over{2i\pi}}(\gamma\psi_m\partial{X^m}
-q\partial\lambda)
\cr
Q_3=-{1\over4}\oint{{dz}\over{2i\pi}}b\gamma^2
}}
Then, in the limit $\mu_0\rightarrow{0}$ the action (2) is symmetric
under the global space-time transformations generated by
\eqn\grav{\eqalign{T^m={1\over{\rho}}
K\circ\oint{dz}e^\phi(\lambda\partial^2{X_m}-2\partial\lambda\partial
{X^m})\cr
T^{mn}=K\circ\oint{dz}\psi^m\psi^n}}
where the homotopy transform of an operator $V$
$K\circ{V}$ is defined according to
\eqn\grav{\eqalign{
K{\circ}V=T+{{(-1)^N}\over{N!}}
\oint{{dz}\over{2i\pi}}(z-w)^N:K\partial^N{W}:(z)
\cr
+{1\over{{N!}}}\oint{{dz}\over{2i\pi}}\partial_z^{N+1}{\lbrack}
(z-w)^N{K}(z)\rbrack{K}\lbrace{Q_{brst}},U\rbrace}}
where $w$ is some arbitrary point on the worldsheet,
$U$ and $W$ are the operators defined according 
to
\eqn\grav{\eqalign{\lbrack{Q_{brst}},V(z)\rbrack=\partial{U}(z)+W(z),}}
\eqn\lowen{K=ce^{2\chi-2\phi}}
is the homotopy operator satisfying
${\lbrace}Q_{brst},K{\rbrace}=1$
and $N$ is the leading order of the operator product
\eqn\lowen{K(z_1)W(z_2)\sim{(z_1-z_2)^N}Y(z_2)+O((z_1-z_2)^{N+1})}
The operators $T^m$ and $T^{mn}$ then can be shown to
satisfy the $AdS$ isometry algebra with the cosmological
constant $\Lambda=-{1\over{\rho^2}}$ ~{\selfpp}
:
\eqn\grav{\eqalign{
{\lbrack}T_{ab},T_{cd}\rbrack
=\eta_{ac}T_{bd}-\eta_{ab}T_{cd}-\eta_{cd}T_{ab}+\eta_{bd}T_{ac}\cr
{\lbrack}T_{a},T_{bc}\rbrack
=\eta_{ab}T_{c}-\eta_{ac}T_{b}\cr
{\lbrack}T_{a},T_{b}\rbrack
=-{1\over{\rho^2}}{T_{ab}}}}
The minus sign in the 
last commutator is actually highly nontrivial and is related
to the  subtleties of OPE and picture equivalence relations
analyzed in ~{\selfpp}. The AdS isometry algebra (11) also
admits another Realization in terms of the operators 
 $(S^m,L^{mn})$
where 
\eqn\lowen{
L^{mn}=K\circ{T^{mn}}\equiv{K}\circ\oint{{dz}\over{2i\pi}}\psi^m\psi^n}
is the same full rotation operator (6) 
(where the $K\circ$ represents the homotopy transformation
to ensure the BRST-invariance)
while $S^{m}$ is the homotopy transformation of the operator
$\oint{{dz}\over{2i\pi}}\lambda\psi^m$, representing
the rotation in the Liouville-matter plane:
\eqn\grav{\eqalign{
S^m=K\circ{\rho^{-1}}\oint{{dz}\over{2i\pi}}\lambda\psi^m
\cr
={{\rho^{-1}}}\oint{{dz}\over{2i\pi}}\lbrack
\lambda\psi^m+2ce^{\chi-\phi}(\partial\varphi\psi^m-\partial{X^m}\lambda
-qP^{(1)}_{\phi-\chi}\psi^m)
-4\partial{c}c{e^{2\chi-2\phi}}\lambda\psi^m\rbrack
\cr
=-4{\lbrace}Q,{\rho^{-1}}
\oint{{dz}\over{2i\pi}}ce^{2\chi-2\phi}\lambda\psi^m\rbrace}}
 
The
conformal weight $n$ polynomials
$P^{(n)}_{a\phi+b\chi+c\sigma}$ (where $a,b,c$ are some constants) are defined
according to
\eqn\lowen{P^{(n)}_{a\phi+b\chi+c\sigma}=e^{-a\phi(z)-b\chi(z)-c\sigma(z)}
{{d^n}\over{dz^n}}e^{a\phi(z)+b\chi(z)+c\sigma(z)}}
(with the product taken in algebraic rather than OPE sense)

Starting from the symmetry generators (6), (13), one can construct
the closed string vertex operator describing 
the dynamics  of vielbeins and spin connection gauge 
fields in space-time ~{\selfpp}

\eqn\grav{\eqalign{
G(p)=e^a_m(p)F_a{\bar{L}}^m+
+\omega^{ab}_m(p)
(F^m_{b}{\bar{L}}_a-{1\over2}F_{ab}{\bar{L}}^m)+c.c.}}
where
\eqn\grav{\eqalign{
F_m=-2K_{U_1}\circ\int{dz}\lambda\psi_me^{ipX}(z)\cr
U_1=\lambda\psi_me^{ipX}+{i\over2}\gamma\lambda
(({\vec{p}}{\vec{\psi}})\psi_m-p_mP^{(1)}_{\phi-\chi})e^{ipX}}}
or manifestly
\eqn\grav{\eqalign{
F_m=-2\int{dz}{\lbrace}\lambda\psi_m(1-4\partial{c}ce^{2\chi-2\phi})+
\cr
2ce^{\chi-\phi}
(\lambda\partial{X}_m-\partial\varphi\psi_m+q\psi_mP^{(1)}_{\phi-\chi}
-{i\over2}(({\vec{p}}{\vec{\psi}})\psi_m-p_mP^{(1)}_{\phi-\chi}))\rbrace{e^{ipX}}(z)}}
where the $partial$ homotopy transform 
$T{\rightarrow}L=K_{\Upsilon}\circ{T}$ of an operator $T$ based on
$\Upsilon$ is defined according to 
\eqn\grav{\eqalign{
{{L}}(w)=K_\Upsilon{\circ}T=T+{{(-1)^N}\over{N!}}
\oint{{dz}\over{2i\pi}}(z-w)^N:K\partial^N{\Upsilon}:(z)
\cr
+{1\over{{N!}}}\oint{{dz}\over{2i\pi}}\partial_z^{N+1}{\lbrack}
(z-w)^N{K}(z)\rbrack{K}\lbrace{Q_{brst}},U\rbrace}}
where $N$ is the leading order of the OPE of $K$ and $\Upsilon$. 
Particularly, if $\lbrack{Q},T\rbrack=\oint\Upsilon$, 
the partial homotopy transform  obviously
coincides with the usual
homotopy transform.
Next,
\eqn\grav{\eqalign{{\bar{L}}^a=\int{d{\bar{z}}}e^{-3{\bar\phi}}
\lbrace{\bar\lambda}\bar\partial^2{X^a}-2\bar\partial\bar\lambda\bar\partial
{X^a}
\cr+
ip^a({1\over2}\bar\partial^2\bar\lambda+{1\over{q}}
\bar\partial\bar\varphi\bar\partial\bar\lambda
-{1\over2}\bar\lambda(\bar\partial\bar\varphi)^2+
(1+3q^2)\bar\lambda(3\bar\partial\bar\psi_b\bar\psi^b-{1\over{2q}}
\bar\partial^2\bar\varphi))\rbrace{e^{ipX}}}}
at the minimal negative picture $-3$ representation 

and 

\eqn\grav{\eqalign{{\bar{L}}^a=K\circ\int{d{\bar{z}}}e^{{\bar\phi}}
\lbrace{\bar\lambda}\bar\partial^2{X^a}-2\bar\partial\bar\lambda
\bar\partial
{X^a}
\cr+
ip^a({1\over2}\bar\partial^2\bar\lambda+{1\over{q}}
\bar\partial\bar\varphi\bar\partial\bar\lambda
-{1\over2}\bar\lambda(\bar\partial\bar\varphi)^2+
(1+3q^2)\bar\lambda(3\bar\partial\bar\psi_b\bar\psi^b-{1\over{2q}}
\bar\partial^2\bar\varphi))\rbrace{e^{ipX}}}}
at the minimal positive picture $+1$ representation.
(similarly for its holomorphic counterpart $L^a$)
Then,
\eqn\grav{\eqalign{
F_{ma}=F_{ma}^{(1)}+F_{ma}^{(2)}+F_{ma}^{(3)}}}
where
\eqn\grav{\eqalign{
F_{ma}^{(1)}=-4qK_{U_2}\circ\int{dz}ce^{\chi-\phi}\lambda\psi_m\psi_a\cr
U_2=\lbrack{Q-Q_3},ce^{\chi-\phi}\lambda\psi_m\psi_a{e^{ipX}}\rbrack
-{i\over2}c\lambda(({\vec{p}}{\vec{\psi}})\psi_a\psi_m-p_m\psi_aP^{(1)}_{\phi-\chi})
e^{ipX}(z)}}
\eqn\grav{\eqalign{
F_{ma}^{(2)}=K\circ\int{dz}\psi_m\psi_a{e^{ipX}}=-4\lbrace{Q},\int{dz}
ce^{2\chi-2\phi}{e^{ipX}}\psi_m\psi_a(z)\rbrace}}
and
\eqn\grav{\eqalign{
F_{ma}^{(3)}=K\circ\int{dz}e^{\phi}(\psi_{\lbrack{m}}\partial^2{X}_{a\rbrack}
-2\partial\psi_{\lbrack{m}}\partial{X}_{a\rbrack})e^{ipX}(z)}}
In the limit  of zero momentum the holomorphic and the antiholomorphic
components of the operator correspond to AdS isometry generators
 in different realizations, described above.
The BRST invariance imposes the following on-shell constraints
on vielbein and connection fields:
\eqn\grav{\eqalign{
p^{\lbrack{n}}e^b_{m\rbrack}(p)-\omega^{b{\lbrack}n}_{m\rbrack}(p)=0\cr
p_{\lbrack{n}}\omega_{m\rbrack}^{ab}(p)=0
\cr
p^m{e_m^b}(p)=0
\cr
p^m\omega_m^{ab}(p)=0}}
The first two constraints represent the linearized
equations $R^{AB}=0$ (the first one being the zero torsion constraint
$T^a=R^{a{\hat{d}}}=0$
while the second reproducing vanishing Lorenz curvature $R^{ab}=0$).
The last two constraints  represent the gauge fixing conditions
related to the diffeomorphism symmetries.
The fact that the BRST invariance leads to space-time equations in
a certain gauge is not surprising if we recall that similar constraints
on a standard vertex operator of a photon also lead to Maxwell's equations in 
the Lorenz gauge.
Provided that the constraints (25) are satisfied, the vertex operator 
$G(p)$
can be written as a BRST commutator in the large Hilbert space
plus terms that are manifestly in the small Hilbert space,
according to
\eqn\grav{\eqalign{
G(p)=\lbrace{Q},W(p)\rbrace+{1\over{q}}K\circ\omega_m^{ab}\int{dz}
e^{\phi}(\psi^{\lbrack{m}}\partial^2{X}_{a\rbrack}
-2\partial\psi_{\lbrack{m}}\partial^{X}_{a\rbrack})e^{ipX}(z)
{\bar{L}}_b+c.c.
\cr
W(p)=8e^a_m(p){\bar{L}}_a\int{dz}
c\partial\xi\xi{e^{-2\phi}}\lambda\psi^me^{ipX}
\cr
+\omega_m^{ab}{\bar{L}}_b{\lbrack}-{4\over{q}}\int{dz}
c\partial\xi\xi{e^{-2\phi}}\psi_a\psi^m{e^{ipX}}
\cr
+4\int{dz}(z-w)\partial{c}c\partial^2\xi\partial\xi\xi{e^{-3\phi}}
\lambda\psi_a\psi^m{e^{ipX}}\rbrack}}
This particularly implies that , modulo gauge transformations,
the vertex operator $G(p)$ is the element of the $small$ Hilbert space.

The linearized gauge symmetry transformations for
vielbein and connection gauge fields are given by:

\eqn\grav{\eqalign{\delta{e_m^a}=\partial_m\rho^{a}+\rho_m^a\cr
\delta\omega_m^{ab}=\partial_m\rho^{ab}+\rho^{\lbrack{a}}\delta^{b\rbrack}_m}}
where we write $\rho^{AB}=(\rho^{ab},\rho^{a{\hat{d}}})=(\rho^{ab},\rho^a)$
The variation of $G(p)$ under (27) in the momentum space is 
\eqn\grav{\eqalign{\delta{G(p)}=
p^mF_m{\bar{L}}_a\rho^a+ p^mF_{ma}{\bar{L}}_b\rho^{ab}}}
The  two terms of the variation (28) are BRST exact in the $small$ 
Hilbert space (and therefore are irrelevant in correlators) 
since
\eqn\grav{\eqalign{
p^mF_m=\lbrace{Q}, :\Gamma:(w)\lbrack{Q},\xi{A}\rbrack\rbrace
\cr
A=\int{dz}e^{\chi-3\phi}\partial\chi(({\vec{p}}{\vec{\partial{X}}})\lambda
-({\vec{p}}{\vec\psi})\partial\varphi+({\vec{p}}{\vec\psi})
P^{(1)}_{\phi-(1+q)\chi}
)e^{ipX}}}

and

\eqn\grav{\eqalign{p^mF_{ma}^{(1)}=
4q\lbrack{Q},\Gamma(w)\int{dz}ce^{-3\phi}\partial\xi\partial^2\xi
\lambda\psi_a({\vec{p}}{\vec{\psi}})e^{ipX}\rbrack\cr
p^mF_{ma}^{(2)}=\lbrace{Q},:\Gamma:(w)
\int{dz}\partial\xi{e^{-3\phi}}(({\vec{p}}{\vec\psi})\partial{X}_a
-({\vec{p}}{\vec{\partial{X}}})\psi_a)e^{ipX}\rbrace\cr
p^mF_{ma}^{(3)}=\lbrace{Q},\lbrack{K}\circ\int{dz}\lambda\psi_a{e^{ipX}},B\rbrack
\rbrace\cr
B=\int{dz}\partial\xi{e^{-4\phi}}\lbrack
\lambda(\partial{\vec\psi}\partial^2{\vec{X}})
-2\partial\lambda(({\vec\psi}\partial^2{\vec{X}})-
2(\partial{\vec\psi}\partial{\vec{X}}))\rbrack
}}

Therefore gauge transformations of $e$ and $\omega$ shift 
$G(p)$ by terms not contributing to correlators.
The $G(p)$ vertex operator, which construction is explained above,
describes the dynamics of spin $s=2$ massless field in the closed
string spectrum, in terms of vielbein and connection gauge fields.
Note that, despite the fact that the unperturbed
theory has been originally defined
in the flat space, in the $perturbed$ theory (which flows
to the AdS vacuum) 
 the distinction betwen the tangent indices $a,b,...$
and the manifold indices $m,n,...$ is ensured
by the corresponding operators of the $F_m$-type
and $L^a$-type being the elements of different ghost cohomologies
and having very different on-shell constraints and gauge symmetries.
In the leading order, the vanishing $beta$-function
condition for the sigma-model,
fiven by the RNS action perturbed by the $G(p)$-operator
(15)
 leads to space-time equations
for $\omega$ and $\lambda$, given by ~{ \selfpp}
\eqn\grav{\eqalign{R^{ab}=d\omega^{ab}
+(\omega\wedge\omega)^{ab}-{1\over{\rho^2}}
e^a\wedge{e^b}=0}}
and
\eqn\lowen{
de^a+\omega^{ab}\wedge{e^b}=0}
with the solution given by the $AdS$ vacuum ~{\mmswf}.
From the $2$-dimensional point of view this means that
the RNS theory, initially defined in the flat space and
perturbed with $G(p)$, flows to new fixed conformal point
corresponding to  theory in the new space-time background,
namely, $AdS$. The next step is to introduce the higher spin 
excitations.
The vertex operators, describing the dynamics
of the massless higher spin $s\geq{3}$ fields in Vasiliev's
frame-like formalism, can be constructed in the $open$ string sector
of the extended RNS superstring theory.Such a construction has been
recently performed in ~{\selfp}.
In the frame-like approach,
the spin 3 field is described by
the dynamical space-time field $\omega^{2|0}\equiv\omega_m^{ab}$, 
as well as by two auxiliary fields
$\omega^{2|1}\equiv\omega_m^{ab|c}$  and $\omega^{2|2}\equiv
\omega_m^{ab|cd}$, related to $\omega^{2|0}$ by generalized
zero torsion constraints ~{\selfp}
 
The vertex operators
for the dynamical  $\omega^{2|0}$-field for the massless 
spin 3 are given by:

\eqn\grav{\eqalign{V^{(-3)}=H_{abm}(p)
c{e^{-3\phi}}\partial{X^a}\partial{X^b}
\psi^me^{ipX}}}
at unintegrated minimal negative picture
and
\eqn\grav{\eqalign{V^{(+1)}=K\circ{H_{abm}(p)}\oint{dz}e^\phi
\partial{X^a}\partial{X^b}
\psi^me^{ipX}}}
at integrated minimal positive picture $+1$.
These operators are the elements of superconformal
ghost cohomology $H_1\sim{H_{-3}}$ ~{\selfp}
The vertex operators for the first auxiliary field
$\omega^{2|1}$ are given by
\eqn\grav{\eqalign{
V_{-}^{2|1}(p)=2\omega_m^{ab|c}(p)ce^{-4\phi}(
-2\partial\psi^m\psi_c\partial{X_{(a}}\partial^2{X_{b)}}
\cr
-2\partial\psi^m\partial\psi_c\partial{X_a}\partial{X_b}
+\psi^{m}\partial^2\psi_{c}\partial{X_a}\partial{X_b})e^{ipX}
}}
at negative (unintegrated) representation
and
\eqn\grav{\eqalign{
V_{+}^{2|1}(p)=2\omega_m^{ab|c}(p)K\circ{\oint{dz}}e^{2\phi}(
-2\partial\psi^m\psi_c\partial{X_{(a}}\partial^2{X_{b)}}
\cr
-2\partial\psi^m\partial\psi_c\partial{X_a}\partial{X_b}
+\psi^{m}\partial^2\psi_{c}\partial{X_a}\partial{X_b})e^{ipX}
}}
at the positive (integrated) representations.
The operators (35), (36) are the elements of $H_2\sim{H_{-4}}$;
the cohomology constraints for $V_{\pm}^{2|1}(p)$
lead to  generalized zero torsion constraints
relating $\omega^{2|1}$ and $\omega^{2|0}$:
\eqn\grav{\eqalign{\omega_m^{ab|c}
=2p^c\omega_m^{ab}-p^a\omega_m^{bc}
}}
modulo BRST exact terms in small Hilbert space.
The vertex operators for the second auxiliary field
$\omega^{2|2}$ are given by
\eqn\grav{\eqalign{
V_{-}^{2|2}(p)=-3\omega_m^{ab|cd}(p)
ce^{-5\phi}(\psi^{m}\partial^2\psi_c\partial
^3\psi_{d}\partial{X^a}\partial{X_b}
-2
\psi^{m}\partial\psi_c\partial
^3\psi_{d}\partial{X_{a}}\partial^2{X_{b}}
\cr
+{5\over8}\psi^{m}\partial\psi_c\partial
^2\psi_{d}\partial{X_{a}}\partial^3{X_{b}}
+{{57}\over{16}}
\psi^{m}\partial\psi_c\partial
^2\psi_{d}\partial^2{X_{a}}\partial^2{X_{b}})e^{ipX}
}}
at negative (unintegrated) representation
and
\eqn\grav{\eqalign{
V_{+}^{2|2}(p)=-3\omega_m^{ab|cd}(p)K\circ{\oint{dz}}
e^{3\phi}(\psi^{m}\partial^2\psi_c\partial
^3\psi_{d}\partial{X^a}\partial{X_b}
-2
\psi^{m}\partial\psi_c\partial
^3\psi_{d}\partial{X_{a}}\partial^2{X_{b}}
\cr
+{5\over8}\psi^{m}\partial\psi_c\partial
^2\psi_{d}\partial{X_{a}}\partial^3{X_{b}}
+{{57}\over{16}}
\psi^{m}\partial\psi_c\partial
^2\psi_{d}\partial^2{X_{a}}\partial^2{X_{b}})e^{ipX}
}}
at positive (integrated) representation.
The $V_\pm^{2|2}$ operators are the elements
of $H_3\sim{H_{-5}}$ and the cohomology constraint
leads to the second generalized torsion condition
relating $\omega^{2|2}(p)$ and $\omega^{2|1}(p)$ up
to BRST exact terms:
\eqn\grav{\eqalign{
\omega_m^{ab|cd}=2p^d\omega^{ab|c}-p^a\omega^{bd|c}
-p^b\omega^{ad|c}+2p^c\omega^{ab|d}-p^a\omega^{bc|d}
-p^b\omega^{ac|d}}}.
Combining the $AdS$ sigma-model construction
~{\selfpp} with expressions for vertex operators
describing the higher spin excitations in unfolded formalism,
the generating functional for the model describing
the higher spin dynamics in AdS space is given by
\eqn\grav{\eqalign{
Z(e_m^a,\omega_m^{ab},\omega^{s-1|t},\rho)
=\int{D}(X,\psi,\bar\psi,ghosts)e^{-S_{RNS}+
\int_{p}{\lbrace}G(p,\rho)
+\sum_s\sum_{t=0}^{s-1}\omega^{s-1|t}(p)V^{s-1|t}(p)\rbrace}}}
where ${\lbrace}\omega^{s-1|t}{\rbrace}$ is the 
set of dynamical and auxiliary fields for the spin $s>2$ and $V^{s-1|t}$
are the corresponding massless vertex operators in open string theory.
In this paper we shall restrict ourselves to the spin $3$ case.
The  correlation functions describing  the higher spin interactions
in the $AdS$ space are then given by
\eqn\grav{\eqalign{
<V^{s_1-1|t_1}(p_1)...V^{s_N-1|t_N}(p_N)>
={{\delta^{(n)}{Z(e_m^a,\omega_m^{ab},\omega^{s-1|t},\rho)}}
\over{\delta{\omega^{s_1-1|t_1}(p_1)}...\delta{\omega^{s_N-1|t_N}(p_N)}}}
|_{\omega^{s_1-1|t_1}=0,...,\omega^{s_N-1|t_N}=0}}}
In the rest of the paper, for the sake of the holographic context,
we shall assume
that all the operators of the $d$-dimensional non-critical superstring
theory are  polarized along the $d-1$-dimensional subspace
and also propagate in this subspace corresponding to the
underlying $AdS_d$  boundary, unless stated otherwise.

Perturbation expansion in the powers of ${1\over{\rho}}$ then describes
the $AdS$ deformations of the higher spin interactions in terms
of $\alpha^\prime$ and the cosmological 
constant $\Lambda$ in the frame-like formalism.
In the next section  we shall use the generating functional
(41) in order to compute the $AdS$ deformations of the $3-vertex$
for the spin 3 fields in the first nontrivial order in $\Lambda$ and 
to  analyze their relevance to the CFT correlators in the dual 
model.

\centerline{\bf 3. Holographic
Spin 3 Vertex in AdS background: preliminaries}

In the previous work ~{\selfp} we computed the three-point function
of the spin 3 vertex operators (33)-(40) in 
open string theory in the flat space, showing it to reproduce the 
Berends-Burgers-Van Dam (BBD) type
of interaction vertex in space-time ~{\bbd} for spin 3 in 
the frame-like
formalism. To compute
the $AdS_d$ deformations of this vertex, 
one has to expand the functional (41) 
in powers of $1\over{\rho}$, that is, $G(p)$. The result 
significantly depends on the number of space-time dimensions since
$G(p)$  expression (15) depends manifestly
on the Liouville background charge. Since $G(p)$ operator for
vielbein and spin 2 connection is a closed string excitation  and
spin 3 fields vertex operators are in the open string spectrum,
the leading order contribution to the $AdS_d$-deformation stems
from the amplitude on the disc. Furthermore , it is clear that
the contribution linear in in the spin 2 connection $\omega_m^{ab}$,
which has the order of $\rho^{-1}\sim{\sqrt{\Lambda}}$ vanishes
since the corresponding correlator is linear in the
Liouville superpartner $\lambda$, i.e. 
is proportional to the vanishing
one-point function of $\lambda$. Similarly, all the contributions
proportional to odd powers of $\rho^{-1}$ or half-integer powers of
$\Lambda$ vanish as well, since they all are given
by the correlators containing odd numbers of $\lambda$ insertions.
For this reason, the first nontrivial leading order contribution
to the disc correlator is proportional to the $AdS_d$ vielbein field
$e_m^a(p)$ and is of the order of $\rho^{-2}$.
This is the contribution given by the 4-point function on the disc,
equivalent to the five-point function on the sphere.
The ghost number selection rule therefore requires that the overall 
left$+$right $\phi$-ghost number  carried by the correlator equals to
$-2$. This selection rule particularly 
makes it convenient to take two spin 3
operators at  the $\omega^{2|0}$ representation and at the negative 
unintegrated ghost picture $-3$ representation (33). 
It is convenient to locate them at the points $z_{1,2}=\pm{i}$ on 
the disc.
The third spin 3 operator is convenient to take at the 
$\omega^{2|2}$-representation (39) and at the minimal positive integrated
ghost picture $+3$ representation. 
There is no loss of generality here, since the operators
for $\omega^{2|0}$ and $\omega^{2|1}$ are the elements of
different ghost cohomologies and do not contribute
due to ghost number selection rules, provided that
  all the $AdS_d$ transvection $L_m$-type
operators in the $G(p)$ insertion are taken at positive $+1$ 
picture representation  while the 
transvection $F_a$-type or the
rotation $F_{ab}$-type operators entering
$G(p)$ are taken at picture $0$ (the latter do not of course
 contribute
to the leading order for the reasons described above).
The $G(p)$ operator is also integrated over the interior of the disc.
Finally, it is convenient to present the manifest form of
the transvection type
operators entering $G(p)$  and of the integrated 
spin 3 operator for $\omega^{2|2}$,
upon applying all the relevant
homotopy/partial homotopy transforms:
\eqn\grav{\eqalign{
F_m=-2\int{dz}{\lbrace}\lambda\psi_m(1-4\partial{c}ce^{2\chi-2\phi})+
\cr
2ce^{\chi-\phi}
(\lambda\partial{X}_m-\partial\varphi\psi_m+q\psi_mP^{(1)}_{\phi-\chi}
-{i\over2}(({\vec{p}}{\vec{\psi}})\psi_m-p_mP^{(1)}_{\phi-\chi}))\rbrace{e^{ipX}}(z)}}
and
\eqn\grav{\eqalign{{{L}}^a(p,u)={1\over2}\int{d{{z}}}(z-u)^2\lbrace
(P^{(2)}_{2\phi-2\chi-\sigma}e^{{\phi}}-24\partial{c}ce^{2\chi-\phi})
\cr\times
\lbrace{\lambda}\partial^2{X^a}-2\partial\lambda\partial
{X^a}
ip^a({1\over2}\partial^2\lambda+{1\over{q}}
\partial\varphi\partial\lambda
-{1\over2}\lambda(\partial\varphi)^2
\cr
+
(1+3q^2)\lambda(3\partial\psi_b\psi^b-{1\over{2q}}
\partial^2\varphi))
+c{e^\chi}G^{(4)}(\phi,\chi,\psi,\lambda,\varphi,X)
\rbrace{e^{ipX}}(z)
}}
where $u$ is an arbitrary point which choice is
irrelevant to the correlators since all the $u$-derivatives
of $L^a$ are BRST-exact in the small Hilbert space ~{\selfc}.
For our purposes, it shall be  particularly convenient to
choose $u=-i$ on the unit disc boundary.
Finally
\eqn\grav{\eqalign{
V_{+}^{2|2}(p)=-3\omega_m^{ab|cd}(p){\oint{dz}}(z-u)^6\lbrace
({1\over{6!}}e^{3\phi}P^{(6)}_{2\phi-2\chi-\sigma}
-28\partial{c}c{e^{2\chi+\phi}})
\cr\times
(\psi^{m}\partial^2\psi_c\partial
^3\psi_{d}\partial{X^a}\partial{X_b}
-2
\psi^{m}\partial\psi_c\partial
^3\psi_{d}\partial{X_{a}}\partial^2{X_{b}}
\cr
+{5\over8}\psi^{m}\partial\psi_c\partial
^2\psi_{d}\partial{X_{a}}\partial^3{X_{b}}
+{{57}\over{16}}
\psi^{m}\partial\psi_c\partial
^2\psi_{d}\partial^2{X_{a}}\partial^2{X_{b}})
\cr
+c{e^{\chi+2\phi}}G^{(12)}(\phi,\chi,\psi,\lambda,\varphi,X)\rbrace
e^{ipX}
}}
where 
$G^{(4)}(\phi,\chi,\psi,\lambda,X)$ and
$G^{(12)}(\phi,\chi,\psi,\lambda,X)$ are 
certain operators of conformal dimensions 4 and 12
accordingly, depending on derivatives of the matter
and Liouville fields $X,\varphi$, 
bosonized ghost fields $\phi$ and $\chi$
and also on $\lambda,\psi$ and their derivatives.
The manifest form of these operators is irrelevant to us
since the pieces proportional to $\sim{c{e^\chi}}$ in $L^{a}$
and to $\sim{c{e^{\chi+2\phi}}}$ in 
$V_{+}^{2|2}(p)$  don't contribute to the overall correlator
 due to the ghost number selection rules.
Similarly, the selection rules exclude the pieces proportional
to $\sim\partial{c}c{e^{2\chi-\phi}}$
 and 
$\sim\partial{c}c{e^{\chi+2\phi}}$
in the expressions (44), (45) for $L^a$ and
$V_{+}^{2|2}(p)$ accordingly. Finally, the selection rules
leave the only relevant term
in the expression (17) for $F_m$ proportional to 
$\sim\oint{dz}\lambda\psi_m$ with all others not contributing to
the leading order correlator for the same reason.
This altogether significantly simplifies the computation
of the 5-point correlator, making it still cumbersome but
not anymore insurmountable.

\centerline{\bf 4. Holographic Spin 3 Vertex in AdS background:
 the computation}
Using the results of the previous section, it is now  straightforward
to identify the correlator giving the $AdS$ deformation of spin 3
vertex in the leading order:
\eqn\grav{\eqalign{
A(p;k_1,k_2,k_3)=e_m^a(p)\omega_{m_3}^{a_3b_3|c_3d_3}(k_1)
\omega_{m_1}^{a_1b_1}(k_2)\omega_{m_2}^{a_2b_2}(k_3)
\cr
\times\lbrace<
\int{d^2z}(z-u)^2{\lbrace}{e^\phi}P^{(2)}_{2\phi-2\chi-\sigma}
(\lambda\partial^2{X_a}-2\partial\lambda\partial{X_a})
\cr
+ip_a({1\over2}\partial^2\lambda+3(1+3q^2)\lambda\partial\psi_b\psi^b)
e^{ipX}(z)\bar\lambda\bar\psi^m{e^{ipX}}({\bar{z}})+c.c.\rbrace
\cr
\int{d\tau}(\tau-\tau_1)^6P^{(6)}_{2\phi-2\chi-\sigma}e^{3\phi}
(\psi^{m}\partial^2\psi_c\partial
^3\psi_{d}\partial{X^a}\partial{X_b}
\cr
-2
\psi^{m}\partial\psi_c\partial
^3\psi_{d}\partial{X_{a}}\partial^2{X_{b}}
\cr
+{5\over8}\psi^{m}\partial\psi_c\partial
^2\psi_{d}\partial{X_{a}}\partial^3{X_{b}}
+{{57}\over{16}}
\psi^{m}\partial\psi_c\partial
^2\psi_{d}\partial^2{X_{a}}\partial^2{X_{b}})e^{ik_1X}(\tau)
\cr
ce^{-3\phi}\psi^{m_1}\partial{X}_{a_1}\partial{X}_{b_1}
e^{ik_2X}(\tau_1)
ce^{-3\phi}\psi^{m_2}\partial{X}_{a_2}\partial{X}_{b_2}
e^{ik_2X}(\tau_2)>\rbrace}}
where $\tau_1=-i$,$\tau_2=i$, the $\tau$-integration is over the disc
boundary and the $z$-integral is over the interior of the disc.
In order to simplify the computations,
the useful strategy is to first perform the conformal transformation
from  the disc to the upper half-plane
using
\eqn\grav{\eqalign{z\rightarrow{i\over2}{{z+i}\over{z-i}}}}.
Then the integrand of the correlator (46)
can be computed on the half-plane and then integrated
 in $\tau$ (which, upon the conformal transformation,
becomes the integral over the real axis). Having done that,
we shall conformally map  the obtained expression back to the disc,
in order to perform the $z$-integration over the disc's interior.
So we start with the first step, that is, computing the integrand
of (46) on the half-plane.

The contributions to this correlator are factorized in terms of
ghost, $\psi-\lambda$ and $X$-parts.
Let us start with the ghost part, given by
\eqn\grav{\eqalign{A_{gh}(\tau,z,\tau_1,\tau_2)=
<e^{3\phi}P^{(6)}(\tau)e^\phi{P^{(2)}}_{2\phi-2\chi-\sigma}(z)ce^{-3\phi}(\tau_1)
ce^{-3\phi}(\tau_2)>}}
Note that, upon the conformal transformation (47) 
we have $\tau_1=0,\tau_2\rightarrow\infty$, so as usual, we only need
the leading order of this correlator in $\tau_2$ (all others
shall result in expressions with negative powers of $\tau_2$ in the
overall correlator, vanishing in the limit $\tau_2\rightarrow\infty$
and corresponding to the pure gauge contributions)
This means that we only should consider
the contractions of the ghost polynomials 
$P^{(2)}_{2\phi-2\chi-\sigma}(z)$ and $P^{(6)}_{2\phi-2\chi-\sigma}(\tau)$
between themselves and with the ghost exponents $e^{3\phi}(\tau)$
$e^{\sigma-3\phi}(\tau_1)$ and $e^\phi(z)$.
First of all, we note  that  (as it is straightforward to check)
the contractions between the ghost polynomials are limited to

\eqn\grav{\eqalign{P^{(2)}_{2\phi-2\chi-\sigma}(z)
P^{(6)}_{2\phi-2\chi-\sigma}(\tau)=
:P^{(2)}_{2\phi-2\chi-\sigma}(z)
P^{(6)}_{2\phi-2\chi-\sigma}(\tau):
\cr
-{{12}\over{(z-\tau)^2}}
P^{(1)}_{2\phi-2\chi-\sigma}(z)
P^{(5)}_{2\phi-2\chi-\sigma}(\tau)}}
Then correlator (48) can be computed using the associate
ghost polynomial (AGP) technique, explained in ~{\selfp}.
The table of the relevant associate ghost polynomials
for $P^{(6)}_{2\phi-2\chi-\sigma}$
is straightforward to compute and is given by (using the
same notations as in ~{\selfp}):
$$\eqalign{
P^{0|6}_{2\phi-2\chi-\sigma|\sigma-3\phi}=0\cr
P^{1|6}_{2\phi-2\chi-\sigma|\sigma-3\phi}={6!}P^{(1)}_{2\phi-2\chi-\sigma}\cr1
P^{2|6}_{2\phi-2\chi-\sigma|\sigma-3\phi}={5\over{2}}\times{6!}P^{(2)}_{2\phi-2\chi-\sigma}}$$
\eqn\grav{\eqalign{
P^{3|6}_{2\phi-2\chi-\sigma|\sigma-3\phi}={5\over{3}}\times{6!}P^{(3)}_{2\phi-2\chi-\sigma}\cr
P^{4|6}_{2\phi-2\chi-\sigma|\sigma-3\phi}={5\over{12}}\times{6!}P^{(4)}_{2\phi-2\chi-\sigma}\cr
P^{5|6}_{2\phi-2\chi-\sigma|\sigma-3\phi}={{211}\over{24}}\times{6!}P^{(5)}_{2\phi-2\chi-\sigma}\cr
P^{6|6}_{2\phi-2\chi-\sigma|\sigma-3\phi}=P^{(6)}_{2\phi-2\chi-\sigma}\cr
P^{0|6}_{2\phi-2\chi-\sigma|\phi}=7!\cr
P^{1|6}_{2\phi-2\chi-\sigma|\phi}=-6\times{6!}P^{(1)}_{2\phi-2\chi-\sigma}\cr
P^{2|6}_{2\phi-2\chi-\sigma|\phi}={{5}\over{2}}\times{6!}P^{(2)}_{2\phi-2\chi-\sigma}\cr
P^{3|6}_{2\phi-2\chi-\sigma|\phi}=-{{2}\over{3}}\times{6!}P^{(3)}_{2\phi-2\chi-\sigma}\cr
P^{4|6}_{2\phi-2\chi-\sigma|\phi}={{1}\over{8}}\times{6!}P^{(4)}_{2\phi-2\chi-\sigma}\cr
P^{5|6}_{2\phi-2\chi-\sigma|\phi}={{5}\over{2}}\times{6!}P^{(5)}_{2\phi-2\chi-\sigma}\cr
P^{6|6}_{2\phi-2\chi-\sigma|\phi}=P^{(6)}_{2\phi-2\chi-\sigma}\cr
P^{0|5}_{2\phi-2\chi-\sigma|\sigma-3\phi}=5!\cr
P^{1|5}_{2\phi-2\chi-\sigma|\sigma-3\phi}=5\times{5!}P^{(1)}_{2\phi-2\chi-\sigma}\cr
P^{2|5}_{2\phi-2\chi-\sigma|\sigma-3\phi}=5\times{5!}P^{(2)}_{2\phi-2\chi-\sigma}\cr
P^{3|5}_{2\phi-2\chi-\sigma|\sigma-3\phi}={{5}\over{3}}\times{5!}P^{(3)}_{2\phi-2\chi-\sigma}\cr
P^{4|5}_{2\phi-2\chi-\sigma|\sigma-3\phi}={{5}\over{24}}
\times{5!}P^{(4)}_{2\phi-2\chi-\sigma}\cr
P^{5|5}_{2\phi-2\chi-\sigma|\sigma-3\phi}=P^{(5)}_{2\phi-2\chi-\sigma}\cr
P^{0|5}_{2\phi-2\chi-\sigma|\phi}=-6!\cr
P^{1|5}_{2\phi-2\chi-\sigma|\phi}=-{{5}\over{2}}\times{6!}
P^{(1)}_{2\phi-2\chi-\sigma}\cr
P^{2|5}_{2\phi-2\chi-\sigma|\phi}=3\times{6!}
P^{(2)}_{2\phi-2\chi-\sigma}\cr
P^{3|5}_{2\phi-2\chi-\sigma|\phi}=-{{19}\over{12}}\times{6!}
P^{(3)}_{2\phi-2\chi-\sigma}\cr
P^{4|5}_{2\phi-2\chi-\sigma|\phi}={{13}\over{24}}\times{6!}
P^{(4)}_{2\phi-2\chi-\sigma}\cr
P^{5|5}_{2\phi-2\chi-\sigma|\phi}=
P^{(5)}_{2\phi-2\chi-\sigma}}}
Using (49) and the table (50) the ghost correlator (48)
 is straightforward
to compute and is given by:
\eqn\grav{\eqalign{{1\over{2!6!}}{lim_{\tau_2\rightarrow\infty}}
A_{gh}(\tau,z,\tau_1,\tau_2)=
\tau_2^4(\tau-\tau_1)^9(z-\tau_1)^3(\tau-z)^{-3}
\cr
\times\lbrace
\lbrack
{{21}\over{(\tau-z)^2}}+
{{10}\over{(\tau_1-z)^2}}
+
{{30}\over{(\tau-z)(z-\tau_1)}}\rbrack
\times
\lbrack
{{7}\over{(\tau-z)^6}}
-{{30}\over{(\tau-z)^5(\tau-\tau_1)}}
\cr
+{{50}\over{(\tau-z)^4(\tau-\tau_1)^2}}
-{{40}\over{(\tau-z)^3(\tau-\tau_1)^3}}
+{{15}\over{(\tau-z)^2(\tau-\tau_1)^4}}
-{{2}\over{(\tau-z)(\tau-\tau_1)^5}}\rbrack
\cr
-{1\over{(\tau-z)^2}}({6\over{\tau-z}}+{5\over{z-\tau_1}})
\times\lbrack
-{6\over{(\tau-z)^5}}-{{75}\over{(\tau-z)^4(\tau-\tau_1)}}
+
{{360}\over{(\tau-z)^3(\tau-\tau_1)^2}}
\cr
-
{{570}\over{(\tau-z)^2(\tau-\tau_1)^3}}
+
{{390}\over{(\tau-z)(\tau-\tau_1)^4}}
+{1\over{(\tau-\tau_1)^5}}\rbrack}}
This concludes the calculation of the ghost factor of the
overall correlator (46).
Next, we shall consider the matter factor of the correlator (46).
Structurally, the $G(p)$ insertion contributes two different matter
  pieces:
the first resulting from the $L_a$-factor of $G(p)$
containing the matter factor proportional to
$L_a^{(1)}\sim{\lambda\partial^2{X_a}-2\partial\lambda\partial{X_a}
+{1\over2}p_a\partial^2\lambda}$ with no $\psi$-dependence
and the second one stemming from the $\psi$-dependent
piece of $L_a$ proportional to
$L_{a}^{(2)}\sim(3+9q^2)p_a\lambda\partial\psi_b\psi^b$.
Second, the matter part of the $V_{+}^{2|2}$ consists of two terms
of the type $\partial^{(M_1)}X_{a_3}\partial^{(M_2)}X_{b_3}
\partial^{(P_1)}\psi^{m_3}\partial^{(P_2)}\psi_{c_3}\partial^{(P_3)}\psi_{d_3}$
with $M_{1,2}$ ranging from 1 to 3, $P_{1,2,3}$ ranging from 0 to 3
and satisfying $M_1+M_2+P_1+P_2+P_3=7$. The structure of the spin 3
interaction is determined by the $\psi$-contractions between
themselves  and by the $X$-contractions between themselves and
with the exponents. The total number of the $X$-fields
in the correlator (46) is equal to 7, so generically, their contractions
with the exponents may bring from 1 to 7 derivatives in the 
cubic vertex. Since the $\omega^{2|2}$ field already contains 
two derivatives, the possible types of $X$-contractions 
result in interaction terms with the number of derivatives
ranging from 3 to 9. The $9$-derivative contribution
with maximum number of derivatives
(corresponding to the case when all the $X$-derivatives
contract to the exponents) is of particular interest to us since,
in the case of $AdS_4$, 
this contribution is related to the  holographic 
correspondence with the $d=3$ vector model correlator of the type
\eqn\grav{\eqalign{
A({\vec{x}}_1,{\vec{x}}_2,{\vec{x}}_3)=
<\Phi_I\partial_{m_1}\partial_{a_1}\partial_{b_1}
\Phi^I({\vec{x}}_1)
\Phi_J\partial_{m_2}\partial_{a_2}\partial_{b_2}\Phi^J({\vec{x}}_2)
\Phi_K\partial_{m_3}\partial_{a_3}\partial_{b_3}\Phi^K
({\vec{x}}_3)>}}
in the dual vector model;
note that
\eqn\grav{\eqalign{
<\Phi_I\Phi^I({\vec{x}}_1)
\Phi_J\Phi^J({\vec{x}}_2)
\Phi_K\Phi^K({\vec{x}}_3)>
\sim{
N|{\vec{x}}_1-{\vec{x}}_2|^{-1}
|{\vec{x}}_1-{\vec{x}}_3|^{-1}
|{\vec{x}}_2-{\vec{x}}_3|^{-1}}.}}

So let us concentrate on the $9$-derivative case first
and on its relevance to the $AdS_4/CFT_3$ duality.
Since
we are interested in  relating the string theory correlator
(46) to the $d=3$ correlator of the type (52)
with the generic set of indices
$m_j,a_j,b_j(j=1,2,3)$, not all of the 
$\psi$-
contractions are actually
relevant to us.
Some of them would result in appearance of the scalar products
of the momenta in the $9$-derivative contribution.
Such terms are of no interest to us since,
in the duality context, they would correspond 
to special degenerate
correlators in $d=3$
where  the polarizations
of $d=3$ operators at ${\vec{x}}_{1,2,3}$ are contracted along
one or more mutual directions.On the other hand, we are interested
to investigate the relevance of the string correlator (46) to  
 the most general form of the $d=3$ correlator (52), i.e. in the 
case with no contractions among  $m_j,a_j,b_j$ indices with
different $j$. Furthermore, we assume that all the indices
are polarized along the $d=3$ boundary of $AdS_4$.
With all these constraints imposed, it is straightforward to check
that the only relevant $9$-derivative contributions to the correlator 
(46)
 stem from the 
the second part of $G(p)$-insertion containing
$L_a^{(2)}$-factor , while the first one, with the $L_a^{(1)}$-factor, only
gives rise to degenerate terms with $m_j,a_j,b_j$-contractions.
The reason is that the $\psi$-correlator pattern for all the terms
involving $L_a^{(1)}$, has the form
\eqn\grav{\eqalign{
lim_{\tau_2\rightarrow\infty}
<\psi^{m_3}\partial^{(p_1)}\psi^{c_3}\partial^{(p_2)}
\psi^{d_3}(\tau){\bar{\psi}}^m({\bar{z}})\psi^{m_1}(\tau_1)\psi^{m_2}(\tau_2)>
\cr
=\tau_2^{-1}(-1)^{p_1+p_2}p_1!p_2!\times{\lbrace}
{{\eta^{m_2m_3}\eta^{d_3m}\eta^{c_3m_1}}\over{(\tau-{\bar{z}})^{p_2+1}
(\tau-\tau_1)^{p_1+1}}}-
{{\eta^{m_2m_3}\eta^{c_3m}\eta^{d_3m_1}}\over{(\tau-{\bar{z}})^{p_1+1}
(\tau-\tau_1)^{p_2+1}}}
{\rbrace}+O(\tau_2^{-2})}}
leading to unwanted degenerate contractions because of the
common$\eta^{m_2m_3}$ factor.
Straightforward computation of the relevant matter
($X+\psi$)-part of the integrand
of (46) involving $L_a^{(2)}$ then gives
$$\eqalign{{lim_{\tau_2\rightarrow\infty}}
A_{matter}(\tau,z,{\bar{z}},\tau_1,\tau_2)=
(3+9q^2)e_m^a(p)\omega_{m_3}^{a_3b_3|c_3d_3}(k_1)
\omega_{m_1}^{a_1b_1}(k_2)\omega_{m_2}^{a_2b_2}(k_3)
\cr
\times\tau_2^{-4}\times\lbrace
12{\lbrack}k_2^{a_3}k_2^{b_3}({1\over{\tau-\tau_1}}-{1\over{\tau-z}}
-{1\over{\tau-{\bar{z}}}})^2+k_3^{a_3}k_3^{b_3}
({1\over{\tau-z}}
+{1\over{\tau-{\bar{z}}}})^2
\cr
-(k_2^{a_3}k_3^{b_3}+k_3^{a_3}k_2^{b_3})
({1\over{\tau-\tau_1}}-{1\over{\tau-z}}
-{1\over{\tau-{\bar{z}}}})
({1\over{\tau-z}}
+{1\over{\tau-{\bar{z}}}})\rbrack
\cr\times
\lbrack{{2\eta^{mm_2}\eta^{cm_1}\eta^{m_3d}}\over{(\tau-\tau_1)^3(\tau-z)^6}}
-{{\eta^{mm_2}\eta^{dm_1}\eta^{m_3c}}\over{(\tau-\tau_1)^4(\tau-z)^5}}
\cr
+
{{4\eta^{mm_3}\eta^{cm_1}\eta^{m_2d}}\over{(\tau-\tau_1)^3(\tau-z)^5
(z-{\bar{z}})}}
-
{{3\eta^{mm_3}\eta^{cm_2}\eta^{m_1d}}\over{(\tau-\tau_1)^4(\tau-z)^4
(\tau-{\bar{z}})}}
\rbrack
\cr
-12
{\lbrack}k_2^{a_3}k_2^{b_3}({1\over{\tau-\tau_1}}
-{1\over{\tau-z}}
-{1\over{\tau-{\bar{z}}}})({1\over{(\tau-\tau_1)^2}}-{1\over{(\tau-z)^2}}
-{1\over{(\tau-{\bar{z}})^2}})
\cr
+k_3^{a_3}k_3^{b_3}
({1\over{\tau-z}}
+{1\over{\tau-{\bar{z}}}})({1\over{(\tau-z)^2}}
+{1\over{(\tau-{\bar{z}})^2}})
\cr
-k_2^{a_3}k_3^{b_3}
({1\over{(\tau-\tau_1)}}-{1\over{(\tau-z)}}
-{1\over{(\tau-{\bar{z}})}})
({1\over{(\tau-z)^2}}
+{1\over{(\tau-{\bar{z}})^2}})
\cr
-k_3^{a_3}k_2^{b_3}
({1\over{(\tau-\tau_1)^2}}-{1\over{(\tau-z)^2}}
-{1\over{(\tau-{\bar{z}})^2}})
({1\over{\tau-z}}
+{1\over{\tau-{\bar{z}}}})\rbrack
\cr\times
\lbrack
{{2\eta^{mm_2}\eta^{cm_1}\eta^{m_3d}}\over{(\tau-\tau_1)^2(\tau-z)^6}}
+
{{4\eta^{mm_3}\eta^{cm_1}\eta^{m_2d}}\over{(\tau-\tau_1)^2(\tau-z)^5
(\tau-{\bar{z}})}}
-
{{2\eta^{mm_3}\eta^{cm_2}\eta^{m_1d}}\over{(\tau-\tau_1)^4(\tau-z)^3
(\tau-{\bar{z}})}}
\rbrack
\cr
+{5\over2}
{\lbrack}k_2^{a_3}k_2^{b_3}({1\over{\tau-\tau_1}}-{1\over{\tau-z}}
-{1\over{\tau-{\bar{z}}}})({1\over{(\tau-\tau_1)^3}}-{1\over{(\tau-z)^3}}
-{1\over{(\tau-{\bar{z}})^3}})
\cr
+k_3^{a_3}k_3^{b_3}
({1\over{\tau-z}}
+{1\over{\tau-{\bar{z}}}})({1\over{(\tau-z)^3}}
+{1\over{(\tau-{\bar{z}})^3}})
\cr
-k_2^{a_3}k_3^{b_3}
({1\over{(\tau-\tau_1)}}-{1\over{(\tau-z)}}
-{1\over{(\tau-{\bar{z}})}})
({1\over{(\tau-z)^3}}
+{1\over{(\tau-{\bar{z}})^3}})
\cr
-k_3^{a_3}k_2^{b_3}
({1\over{(\tau-\tau_1)^3}}-{1\over{(\tau-z)^3}}
-{1\over{(\tau-{\bar{z}})^2}})
({1\over{\tau-z}}
+{1\over{\tau-{\bar{z}}}})\rbrack
\cr\times
\lbrack
{{\eta^{mm_2}\eta^{cm_1}\eta^{m_3d}}\over{(\tau-\tau_1)^2(\tau-z)^5}}
+
{{3\eta^{mm_3}\eta^{cm_1}\eta^{m_2d}}\over{(\tau-\tau_1)^2(\tau-z)^4
(\tau-{\bar{z}})}}
-
{{2\eta^{mm_3}\eta^{cm_2}\eta^{m_1d}}\over{(\tau-\tau_1)^3(\tau-z)^3
(\tau-{\bar{z}})}}
\rbrack
\cr
+{{57}\over{8}}
{\lbrack}k_2^{a_3}k_2^{b_3}({1\over{(\tau-\tau_1)^2}}-{1\over{(\tau-z)^2}}
-{1\over{(\tau-{\bar{z}})^2}})^2
\cr
+k_3^{a_3}k_3^{b_3}
({1\over{(\tau-z)^2}}
+{1\over{(\tau-{\bar{z}})^2}})^2
\cr
+(k_2^{a_3}k_3^{b_3}+k_3^{a_3}k_2^{b_3})
({1\over{(\tau-\tau_1)^2}}-{1\over{(\tau-z)^2}}
-{1\over{(\tau-{\bar{z}})^2}})
({1\over{(\tau-z)^2}}
+{1\over{(\tau-{\bar{z}})^2}})\rbrack
}$$
\eqn\grav{\eqalign{
\cr\times
\lbrack{{\eta^{mm_2}\eta^{cm_1}\eta^{m_3d}}\over{(\tau-\tau_1)^2(\tau-z)^5}}
+{{3\eta^{mm_3}\eta^{dm_2}\eta^{m_1c}}\over{(\tau-\tau_1)^2(\tau-z)^4
(\tau-{\bar{z}})}}
-
{{2\eta^{mm_3}\eta^{cm_2}\eta^{m_1d}}\over{(\tau-\tau_1)^2(\tau-z)^4
(\tau-{\bar{z}})}}
\rbrack
\rbrace
\cr
{\times}
{\lbrace}{\lbrack}k_1^{a_1}k_1^{b_1}({1\over{\tau-\tau_1}}+{1\over{\tau_1-z}}
+{1\over{\tau_1-{\bar{z}}}})^2
+k_3^{a_1}k_3^{b_1}
({1\over{\tau_1-z}}
+{1\over{\tau_1-{\bar{z}}}})^2
\cr
+(k_1^{a_1}k_3^{b_1}+k_3^{a_1}k_1^{b_1})
({1\over{\tau-\tau_1}}+{1\over{\tau_1-z}}
+{1\over{\tau_1-{\bar{z}}}})
({1\over{\tau_1-z}}
+{1\over{\tau_1-{\bar{z}}}})\rbrack
\cr
\times
{\lbrack}
k_1^{a_2}k_1^{b_2}+k_2^{a_2}k_2^{b_2}+k_1^{a_2}k_2^{b_2}+k_1^{b_2}k_1^{a_2}{\rbrack}
\times
{{(k_1^a+k_2^a+k_3^a)}\over{z-{\bar{z}}}}\rbrace
\cr
\times
|\tau-z|^{-2k_1k_2-2k_1k_3}(\tau-\tau_1)^{k_1k_2}|\tau_1-z|^{-2k_1k_2-2k_2k_3}
+O(\tau_2^{-5})
}}
where we used the momentum conservation 
along with the on-shell conditions on the space-time fields.
The next step is to perform the  integrations in $\tau$ and $z$.
We start with the integral over $\tau$ which, upon the conformal
transformation (47), is along the real line.
As was mentioned above, a 
 convenient choice for $\tau_1$ is $\tau_1=-i$ on the disc
corresponding to $\tau_1=0$ on the half-plane.
The overall integral is given by
\eqn\grav{\eqalign{
A(p;k_1,k_2,k_3)={1\over{2!6!}}\int_{-\infty}^{\infty}d\tau\tau^6
\int{d^2z}z^2{\lbrace}
A_{matter}(\tau,z,{\bar{z}})
A_{gh}(\tau,z)\rbrace}}
where
\eqn\grav{\eqalign{
A_{matter}(\tau,z,{\bar{z}})
A_{gh}(\tau,z)
\equiv
A_{matter}(\tau,z,{\bar{z}},\tau_1,\tau_2)
A_{gh}(\tau,z,\tau_1,\tau_2)|_{\tau_1={0},\tau_2\rightarrow\infty}}}
and we used the fact that
the  leading order
$\sim\tau_2^{4}$-factor of the ghost part of the correlator is cancelled
by the leading order $\sim\tau_2^{-4}$-factor of its matter part.
The $\tau$ integral in (46) looks tricky to evaluate.
However, for our purposes we only need its asymptotic value in the field
theory limit, that is, in the leading order of $\alpha^\prime$.
In this limit the $\tau$ integral  is dominated by 
contributions from the region $\tau\sim{z}$  as the integrand becomes
singular when
 $z$ approaches the 
real axis. In this case, we shall use the asymptotic formula
$$\eqalign{lim_{\epsilon\rightarrow{0}}
\int{d\tau}\int{d^2z}f(z,{\bar{z}})g(\tau,{\bar{z}})
(\tau-z)^{\epsilon-N}
\cr
\sim{{(-1)^{N-1}}\over{(N-1)!\epsilon}}
{\lbrace}\int{d^2z}f(z,{\bar{z}})\partial^{(N-1)}g(z,{\bar{z}})
+O(\epsilon){\rbrace}}$$
(where $\epsilon\equiv{-k_1k_2-k_1k_3}=k_2k_3$)
The result (which is valid up to subleading $\alpha^\prime$-corrections)
is given by the lengthy expression
(function in $z$ and ${\bar{z}}$)
 presented in the Appendix.
Finally, it remains to evaluate  the $z$-integral of (62), (63).
The integrand of (63) is cumbersome  but
structurally all  of the terms are of the type:
\eqn\grav{\eqalign{I(k_1,k_2,k_3)\sim
{(k_2k_3)^{-1}}
{\int}d^2z{\lbrace}{z^{-k_2k_3+N_1}{\bar{z}}^{-k_1k_2-k_1k_3+N_2}
(z-{\bar{z}})^{-k_1k_2-k_1k_3-N_3}+c.c.{\rbrace}}}}
where $N_{1,2,3}$ are some integer numbers, different
for each of the terms entering (63).
The integrals of the type (58) are over the upper half-plane and 
are still
tedious to evaluate.It is therefore convenient, by
using the overall conformal
invariance of  the overall correlator (46) to conformally map 
it back to the unit disc $(z,{\bar{z}})\rightarrow(u,{\bar{u}})$
and introducing $u=re^{i\alpha}$
for the disc coordinates.Then, the transformation
(47)  reduces the integrals of the type (58)
 to those of the generalized elliptic type:
\eqn\grav{\eqalign{I(k_1,k_2,k_3){\sim}2^{-N_1-N_2}
\int_0^1{drr}({{(r^2-1)}\over{(r^2+1)}})^{N_1+N_2-N_3
-2k_1k_2-2k_1k_3-2k_2k_3}
\cr\times
\int_0^{2\pi}d\alpha
(1+{{2r}\over{r^2-1}}cos\alpha)^{-k_1k_3-k_2k_3+N_1}
(1-{{2r}\over{r^2-1}}cos\alpha)^{-k_1k_2-k_2k_3+N_2}
\cr
+(N_1\leftrightarrow{N_2},k_2\leftrightarrow{k_3})}}
The overall amplitude (46) is then given by the lengthy 
expression (63)
described in the Appendix; the answer, however, simplifies
 in the field theory limit of $\alpha^\prime\rightarrow{0}$
Integrating the amplitude (46) over $k_1,k_2,k_3$ and $p$,
using the momentum conservation in the $5$-point amplitude
that eliminates the integral over $p$, 
and recovering the $\alpha^{\prime}$ and the cosmological
constant factors,
the asymptotics
 of  (46) in the field theory limit gives:
\eqn\grav{\eqalign{
A(k_1,k_2,k_3)
\cr
={{691072283467i}\over{360}}\alpha^{\prime}\Lambda
(k_1k_2)^{-1}
(k_1k_3)^{-1}
(k_2k_3)^{-1}
\omega_{m_1}^{a_1b_1}(k_1)\omega_{m_2}^{a_2b_2}(k_2)\omega_{m_3}^{a_3b_3}(k_3)
\cr\times
\lbrace
k_2^{m_1}(k_2)_{a_1}(k_2)_{b_1}
(k_3)^{m_2}(k_3)_{a_2}(k_3)_{b_2}(k_1)^{m_3}(k_1)_{a_3}(k_3)_{b_2}
\cr
+
k_3^{m_1}(k_3)_{a_1}(k_3)_{b_1}
(k_3)^{m_2}(k_3)_{a_2}(k_3)_{b_2}(k_2)^{m_3}(k_2)_{a_3}(k_2)_{b_2}
\rbrace+...}}

where we skipped the contact terms (proportional
to the delta-functions in the position space),
used the zero torsion conditions (37), (40) relating $\omega^{2|2}$
to $\omega^{2|0}$ along with the on-shell conditions for $\omega$'s
and neglected the contributions in
 the subleading order in $\alpha^\prime$.
The overall numerical factor 
in 
(60) is
 consistent with the one obtained in the three-point
string amplitude of spin 3 particles in flat space,
computed in ~{\selfp}; it can therefore be 
be absorbed by the appropriate rescaling
of the vertex operators, making the normalization
obtained from string theory, consistent with the
one in the BBD vertex ~{\boulanger}.
Then the normalization of (60) is consistent
with the one in (52),(53) provided that one identifies
$(\alpha^\prime\Lambda)^2=N^{-1}$ and rescales the vector
field according to 
$\Phi^I\rightarrow{N^{-1\over{4}}}\Phi^I$, which normalizes
 the current's
two-point correlator by 1.
This , up to contact terms (proportional
to the delta-functions in the position space), coincides
with the correlator (52) transformed to the momentum space.

Next, evaluating the lower derivative terms
in the amplitude (56) gives
the answer 
proportional to
\eqn\grav{\eqalign{
A_{lower-der.}(k_1,k_2,k_3)
={(15-6q^2)I_{lower-der}(k_1,k_2,k_3)
+...}}}
with the contact terms skipped.
The explicit expression for $I_{lower-der}(k_1,k_2,k_3)$) 
is given in the Appendix; (61) particularly implies
that, apart from the contact terms,
 the only lower derivative contribution to the overall amplitude
(46)  is proportional to the factor of $\sim{15-6q^2}$ which stems
from $q$-independent and $q$-dependent 
$L_a^{(1)}$ and $L_a^{(2)}$ pieces of the closed string
insertion for the vielbein vertex operator (satisfying
the $AdS$ vacuum solution in the leading order of the beta-function).
This expression vanishes
(up to contact terms and those of the higher order in $\alpha^\prime$)
in $d=4$  where $q={\sqrt{{9-d}\over2}}={\sqrt{{5\over2}}}$.
This means that in the special case of $d=4$ only the nine-derivative
contribution survives in the string-theoretic amplitude (46)
which gives precise holographic relation between AdS string sigma-model
(41) in the case of $d=4$
 and the dual free field theory correlator (53) in $d=3$ for  spin 3.
Thus the $AdS_4/CFT_3$ holography correspondence  for higher spins
appears to be
 surprisingly related to the value of the Liouville background
charge value in $d=4$ which stems from two-dimensional CFT.
This fact is by itself quite intriguing and definitely
needs further investigation.

\centerline{\bf 5. Conclusion and Discussion}

In this paper we analyzed the $AdS_4/CFT_3$ higher spin holography
using string-theoretic sigma-model describing gravity and higher
spin perturbations  around $AdS$ background in the low-energy limit.
We  found that, in the leading order in $\alpha^\prime$ and cosmological 
constant, the three-point correlator for spin 3 particles in $AdS_4$
reproduces the free-field correlator for the large $N$ vector model in 
$d=3$. Surprisingly, we found that this holographic correspondence
appears to be related to the value of the Liouville background
charge in $d=4$ which allows for cancellation of the lower-derivative
terms (up to contact terms),
so the leading order of $AdS$ string theory appears
to be in agreement with Maldacena-Zhiboedov's proposal 
~{\maldaz, \maldazh}. 
 One definitely needs to check whether
such a cancellation also holds for vertex operators 
for spins greater
than 4 and for higher point correlators in the $AdS_4$ case.
On the other hand , the lower 
derivative terms do persist in the three-point
amplitudes for $d\neq{4}$. This is the
signal that the higher spin / CFT holography  has more complicated
character in higher  dimensions, where the dual
theories are no longer free. Moreover, 
even in $AdS_4$ the string theory 
corrections may definitely modify the limit in which 
Maldacena-Zhiboedov's theorem holds. We hope to implement these
computations in the future papers.
The results of this paper suggest 
that string theoretic approach may 
provide interesting insights to HS/CFT duality, 
such as the relevance
of the Liouville theory to the $d=4$ case.
It would be interesting to see possible relations of this fact 
to the $AGT$ conjecture
  since open string
amplitudes for spin $1$ in the sigma-model (41) should particularly
involve the  super Yang-Mills theory in the low energy limit.
Another question of immediate interest 
is to use the sigma-model (41)
in order to study the $AdS_5/CFT_4$ holography for higher spins.
To  approach this problem, one has to study the lower derivative
terms appearing in the sigma-model correlators, 
as well as the higher
order corrections in the cosmological constant. Finally,
in the $AdS_4/CFT_4$ case it would be
of crucial importance to unterstand the relation between
string-theory formalism and 
the twistor space approach used by Vasiliev
~{\hvs} to study the higher spin holography. This relation may
probably, in some form or another, involve the modifications of
 twistor string theory developed by Witten ~{\wittwist}
 This altogether gives
the list of problems to address in the future which of course
is still  very preliminary and incomplete.

\centerline{\bf Acknowledgements}

This work has been supported in part by 
the National Research Foundation (NRF) of Republic of Korea
under the Project no. 2012-004581.
I wish to express my gratitude to the organizers
of the  Vienna Workshop on Higher Spin Gravity at the Erwin
Schroedinger Institute (ESI) during April 10-20 2012, where
part of this work was completed.
I also would like to acknowledge useful
discussions with  S.-J. Rey and M. Vasiliev.

\vfill\eject

\centerline{\bf Appendix}

In this section we present explicit 
expressions for the amplitude (46)
leading to the asymptotics (60). To abbreviate the expressions,
we adopt the following notations:

\eqn\grav{\eqalign{
a\equiv{z}\cr
b\equiv{\bar{z}}\cr
c\equiv{k_1^{a_1}k_1^{b_1}}\cr
d\equiv{k_1^{a_1}k_3^{b_1}+k_1^{b_1}k_3^{a_1}}\cr
f\equiv{k_2^{a_3}k_2^{b_3}}\cr
g\equiv{k_2^{a_3}k_3^{b_3}+k_3^{a_3}k_2^{b_3}}\cr
h\equiv{k_3^{a_3}k_3^{b_3}}\cr
k\equiv{k_3^{a_1}k_3^{b_1}}\cr
p\equiv\eta^{mm_2}\eta^{cm_1}\eta^{m_3d}\cr
q\equiv\eta^{mm_3}\eta^{cm_1}\eta^{m_2d}\cr
t\equiv\eta^{mm_3}\eta^{cm_2}\eta^{m_1d}\cr
u\equiv\omega_{m_1}^{a_1b_1}(k_2)\omega_{m_2}^{a_2b_2}(k_3)
\omega_{m_3}^{a_3b_3|cd}(k_1)
\cr\times
(k_1^a+k_2^a+k_3^a)
\cr\times
(k_1^{a_2}k_1^{b_2}+k_2^{a_2}k_1^{b_2}
+k_1^{a_2}k_2^{b_2}+k_1^{b_2}k_1^{a_2})\cr
L_1=k_1k_2+k_1k_3\cr
L_2=k_1k_2+k_2k_3\cr
L_3=k_1k_3+k_2k_3
}}
Then the amplitude can be expressed covariantly
in a convenient way, suppressing the indices,
in terms of the variables $a,b,c,d,f,g,h,k,q,t,u,L_{1,2,3}$.
The evaluation of the  $\tau$ integral then leads
to the following answer:
$$\eqalign{
A(k_1,k_2,k_3)=
u\int{d{a}d{b}}
\lbrace{{fa^{L_3}b^{L_2}}\over{16(a-b)^{19+L_1}}}
\cr
{\times}{\lbrack}a b(
-8 a^{13} (b (3592376k + 203232 c - 470026 d) -
45 (1488k + 122 c - 231 d)) t +
\cr
24 a^{14} (1488k + 122 c - 231 d) t +
10 b^{13} (4k + c - d) ( - 576 (-20 + 13 b) p 
\cr
+
38925 q - 29952 b q - 66270 t + 32448 b t) 
+
a^9 b^4 (2 (b (-27583068 k + 2090940 c - 780802 d) 
\cr
-
15 (-71416 k + 6175 c + 249 d))  -
384 ( (-53420 k + 736 c - 491 d)) p - 2414880 k q +
\cr
89547132 b q + 595980 c q - 13527948 b c q - 188055 d q +
8919888 b d q - 38145120 k t  
\cr
+ 442112208 b k t - 6170220 c t +
40507190 b c t + 4309920 d t - 111451878 b d t) +
\cr
a^6 b^7 ((b (36984432 k + 5216119 c - 22448132 d) 
-
30 (952240 k + 55785 c - 179903 d)) -
\cr
384 (b (-31790 k + 873 c - 13993 d) + 105 (-16 k + 5 c + 26 d)) p +
\cr
79359120 k q - 51353040 b k q + 4399245 c q - 7162686 b c q 
\cr
-
12648735 d q + 34253916 b d q - 234435600 k t + 204418288 b k t 
\cr
+
8927160 c t - 48353746 b c t + 26078880 d t - 72924756 b d t) 
\cr
+
a^{10} b^3 (2 (b (37160040 k + 1092363 c - 3283276 d) -
\cr
15 (114032 k + 3351 c - 10653 d))-
192 (b(36512k+782c-2099d)-30(80k+c-4d)) p +
\cr
10582200 k q - 233099616 b k q + 302985 c q - 6588324 b c q -
976095 d q + 20179833 b d q 
\cr
- 102815400 k t + 1209791888 b k t -
1626570 c t + 30153268 b c t + 7224210 d t - 59663780 b d t)
\cr
 +
a^3 b^{10} ((2 b (21928972 k + 8677069 c - 7234681 d) -
\cr
15 (1299632 k + 41244 c - 216481 d)) -
192 (-22800 k + 19668 b k - 3540 c 
\cr
+ 34411 b c + 6495 d -
19339 b d) p + 31623120 k q - 63763176 b q + 2049300 c q -
\cr
28605954 b c q - 5653440 d q + 22235154 b d q - 29434440 k t +
\cr
37177976 b k t - 1858710 c t + 40276674 b c t + 6241860 d t -
23223764 b d t) 
\cr
-
10 a^2 b^{11} ((b (2054748 k + 414913 c - 470605 d) +
\cr
12 (-4456 k + 17251 c - 7576 d))  -
96 (-2688 k + 8332 b + 348 c 
\cr
+ 2525 b c + 252 d - 2369 b d) p +
203220 k q - 3367224 b k q 
\cr
- 243405 c q - 865908 b c q +
74475 d q + 871407 b d q - 695820 k t 
\cr
+ 3149708 b k t +
203355 c t + 1343541 b c t + 57525 d t - 1109229 b d t)
}$$
\vfill\eject
$$\eqalign{
 -
10 a b^{12} ((-30 (12076 k + 3989 c - 3504 d) +
\cr
13 b (1044 k + 2573 c - 1417 d)) +
192 (13 b (60 k + 7 c - 11 d) 
\cr
+ 75 (4 k + 3 c - 2 d)) p +
355680 k q + 391248 b k q + 169920 c q 
\cr
+ 30342 b c q - 129420 d q -
64077 b d q + 135120 k t - 1255384 b k t - 168120 c t -
\cr
198978 b c t + 67170 d t + 256412 b d t) 
\cr
+
a^{11} b^2 ((b (9576896 k + 246342 c - 898206 d) -
\cr
15 (15472 k + 542 c - 1683 d)) +
192 (2 b (c + 194 (-20 k + d)) - 15 (-16 k + d)) p 
\cr
+ 696240 k q -
28709400 b k q + 24390 c q - 738933 b c q - 75735 d q 
\cr
+
2693463 b d q + 18384600 k t - 181700576 b k t + 205590 c t 
\cr
-
11289448 b c t - 556680 d t + 6229332 b d t) 
+
a^4 b^9 ((30 (438608 k + 139259 c - 170889 d) 
\cr
+
b (31195608 k - 23625295 c + 7077692 d)) 
192 (30 (496 k - 245 c + 130 d) 
\cr
+
b (-109400 k + 42721 c + 1074 d)) p - 34454160 k q 
\cr
-
50747856 b k q - 8345745 c q + 36016704 b c q + 10265985 d q 
-
9433944 b d q 
\cr
+ 73283520 k t + 39522728 b k t + 15467220 c t -
75827612 b c t - 19635420 d t + 22658110 b d t) 
\cr
+
a^5 b^8 ((-30 (-993296 k + 65495 c + 51431 d) +
b (-140574840 k + 8076197 c + 20592266 d))
\cr
 -
192 (-420 (-100 k + 11 c + 11 d) +
b (-154056 k + 21149 c + 35272 d)) p - 53598240k q 
\cr
+
278165748 b k q + 3934800 c q - 4726128 b c q + 2016675 d q -
42460149 b d q 
\cr
+ 63153600 t - 439526024 b t - 21984180 c t +
63570616 b c t + 8993340 d t + 52194190 b d t)
\cr
+
a^8 b^5 ((30 (632960 k + 25861 c - 55751 d) +
b (-155995128 k - 11490326 c + 21567689 d))
\cr
+
192 (b (42216 k + 6692 c - 19703 d) 
- 30 (320 k + 23 c - 34 d)) p -
55011960 q 
\cr
+ 441897552 b q - 2069865 c q + 30049308 b c q +
4492845 d q - 53982408 b d q + 247050960 k t 
\cr
- 1075703328 b k t +
12480240 c t - 16542330 b c t - 12267000 d t + 64259624 b d t) 
\cr
+
a^7 b^6 ((30 (-527920 k + 21473 c - 22362 d) +
b (109033376 - 1998049 c + 328756 d))  
\cr
-
192 (b (225668 k + 1180 c - 13133 d) -
30 (928 k + 22 c + 13 d)) p 
\cr
+ 27250560 k q - 132110208 b k q -
2015190 c q + 4628397 b c q + 2606355 d q 
- 9431328 b d q 
\cr
-
13143120 k t + 158121584 b k t - 5041440 c t + 2368354 b c t -
18233580 d t + 75056032 b d t)
\cr
+
a^{12}b (120 (105076 k + 4767 c - 12401 d) t +
b ( - 192 (-16 + d) p + 46416 k q}
$$
\vfill\eject
$$\eqalign{
+
1626 c q - 5049 d q - 347866128 k t - 12118888 c t +
34908536 d t)))
\cr
+\lbrace{{ha^{L_3}b^{L_2}}\over{8(a-b)^{19+L_1}}}
{\times}
\cr
{\lbrack}
a b (4 a^{14} (8462 k + 577 c - 1202 d) t +
2 a^{13} (30 (8462 k + 577 c - 1202 d) 
\cr
+
b (-12742604 k - 632218 c + 1559205 d)) t +
30 b^{13} (4 k + c - d) ((960 - 416 b) k p 
\cr
+
7005 k q - 4108 b k q - 7260 k t + 2106 b k t) +
a^9 b^4 (2 (4 b (-1601611k + 615629 c - 659330 d) 
\cr
+
15 (4848 k - 7679 c + 6613 d)) -
192 (-90 c + b (-77608 k + 1129 c - 646 d) 
\cr
+ 60 (81k + d)) p -
348480 k q + 36705972 b k q + 686340 c q - 14675466 b c q 
\cr
-
583785 d q + 15626610 b d q - 23306640 k t + 389991016 b k t -
3838620 c t + 22950092 b c t 
\cr
+ 3457260 d t - 73172098 b d t) +
\cr
a^{10} b^3 ((b (79470528 k + 2016958 c - 6521170 d) -
30 (123808 + 3051 c - 10481 d)) 
\cr
-
96 (b (40888 k + 780 c - 2883 d) - 30 (100 k + c - 5 d)) p +
\cr
11146320 k q - 238310688 b k q + 274005 c q - 6035748 b c q -
942075 d q + 19516749 b d q 
\cr
- 77125680 k t + 841941584 b k t -
1654440 c t + 26608962 b c t + 5691120 d t - 50462452 b d t) 
\cr
+
a^6 b^7 ((b (23733672 k + 3102799 c - 8854348 d) -
30 (889920 k + 38513 c - 122707 d)) 
\cr
-
384 (b (-46491 k + 10514 c - 1893 d) + 105 (28 k - 5 c + 13 d)) p +
76062960 k q 
\cr
- 42926256 b k q + 3536955 c q - 9361710 b c q -
10353825 d q + 17679426 b d q - 139153920 k t 
\cr
+ 36775152 b k t +
875940 c t - 7023550 b c t + 12947220 d t - 5302270 b d t) 
\cr
+
30 a b^{12} ((13 b (3084 k + 355 c - 563 d) +
160 (44k + 57 c - 34 d))
\cr
-
32 (45 (-4k + c) + 13 b (60k + 11 c - 13 d)) p + 44880 k q 
\cr
-
154960 b k q - 7500 c q - 26598 b c q - 1860 d q + 32669 b d q 
\cr
-
193440 kt + 246376 b kt - 19440 c t + 49010 b c t + 33900 d t -
55302 b d t) 
\cr
+
a^4 b^9 ((30 (443488k + 70935 c - 97085 d) +
b (-1151176k - 7947881 c + 3618600 d)) 
\cr
-
96 (b (143928k + 7821 c - 28474 d) +
60 (-512k + 133 c - 2 d)) p - 35452800 q 
\cr
+ 1654464 b q -
5650695 c q + 11720508 b c q + 7261695 d q - 4865790 b d q 
\cr
+
44217240 k t - 31887760 b k t + 4031520 c t + 100832 b c t -
5912910 d t + 247636 b d t) 
\cr
+
a^{11} b^2 ((b (9997696 k + 207342 c - 832814 d) -
1155 (224k + 6 c - 21 d)) 
\cr
+
96 (-15 (-24k + d) + 2 b (-5444k + c + 193 d)) p + 776160 kq 
}$$
\vfill\eject
$$\eqalign{
-
29992848 b kq + 20790 c q - 622065 b c q - 72765 d q +
2498523 b d q 
\cr
+ 11616720 t - 103495104 b t + 249960 c t -
9790670 b c t - 410820 d t + 6061460 b d t) 
\cr
+
a^5 b^8 ((30 (314384k - 51493 c + 13147 d) +
b (-69230024k - 155149 c + 9845590 d))
\cr
+
96 (210 (7 c + 40 (-5k + d)) +
b (85184k + 39575 c - 45454 d)) p - 24069600 k q 
\cr
+
172712364 b k q + 3804840 c q + 9101964 b c q - 925515 d q -
26732397 b d q 
\cr
- 1771320 k t - 129295912 b k t - 7515360 c t -
9836246 b c t + 6549780 d t + 19131876 b d t) 
\cr
+
a^8 b^5 ((30 (606368k + 18269 c - 47855 d) +
b (-149759256k - 7429006 c + 16898695 d)) 
\cr
+
96 (b (-21696k + 7168 c - 32107 d) - 300 (20k + 2 c - 7 d)) p 
\cr
-
53472600 kq + 431464896 b kq - 1644975 c q + 22336260 b c q +
4232835 d q - 49353342 b d q 
\cr
+ 162720240 kt - 573713128 b kt +
7605180 c t + 2790792 b c t 
- 10584600 d t + 27614802 b d t) 
\cr
+
a^{12} b (30 (357076k + 15242 c - 40239 d) t +
b (- 96 (-24k + d) p + 51744 k q 
\cr
+
1386 c q - 4851 d q - 283508960 k t - 9804142 c t +
27924468 d t)) 
\cr
-
2 a^2 b^{11} ( -
3 (16 (b (21044k + 8251 c - 6951 d) -
30 (572k + 83 c - 123 d)) p 
\cr
+ (13 b (96672k + 48818 c -
38053 d) 
+ 480 (304k + 51 c + 4 d)) q 
\cr
+
2 (5 (-8316k + 27871 c - 19191 d) +
b (842212k - 186857 c + 3817 d)) t)) 
\cr
+
a^3 b^{10} (
-2 (48 (b (-35828k + 11269 c - 571 d) -
75 (184k + 106 c - 89 d)) p 
\cr
+
3 (b (3524768k + 1583210 c - 1123521 d) -
60 (29492k + 3877 c - 5475 d)) q 
\cr
+
2 (b (7128476k + 410553 c - 1514936 d) +
45 (-57308k + 3915 c + 6786 d)) t)) 
\cr
+
a^7 b^6 (
3 (32 (-30 (-1192k + 17 c + 53 d) +
b (-257540k + 10748 c + 41515 d)) p 
\cr
+ (b (1784512k +
4673403 c - 6328878 d) +
15 (259520k - 56778 c + 80197 d)) q 
\cr
-
2 (b (18432028k + 2063821 c - 6743278 d) +
10 (-140688k + 6877 c + 155327 d)) t)))
\rbrack\cr
+
\lbrack{{ga^{L_3-3}b^{L_2}}\over{8(a-b)^{19+L_1}}}
((96 a^{20} + 2 a^3 (4835 - 8208 b) b^{16} 
\cr
+
2 a (95 - 48 b) b^{18} - 10 b^{19} + 114 a^2 b^{17} (-15 + 16 b) -
2 a^{19} (5 + 912 b) 
\cr
+
2 a^{18} b )( 3248k - 371 c + 575 d) t +
a^{16} b^2 (- 30 (102632k + 6640 c - 14211 d) t 
\cr
+
b ( 96 (-20k + d) p -
37500 k q - 1116 c q + 3735 d q + 101842384 k t 
\cr
+ 4629100 c t -
11840166 d t))) 
\cr
+
a^{10} b^8 (
b (
96 (-124772k + 21901 c - 17779 d) p 
}$$
\vfill\eject
$$\eqalign{
+ 69337080 k q -
25027488 c q + 22411995 d q + 47573000 k t + 30096024 c t 
\cr
-
71921924 d t) -
15 (
672 (-40 k + 5 c - 52 d) p + 4659444 k q + 5988 c q 
\cr
-
137073 d q - 2235256 k t - 50576 c t - 607218 d t))
\cr
 +
2 a^4 b^{14} (
150 (4k + c - d) (- 9 (16 p + 79 q - 64 t)) 
\cr
-
5 b ( - 2496 (4k + c - d) p - 26988 kq -
6747 c q + 6747 d q + 7280 kt + 1820 c t - 1820 d t))
\cr
 +
a^{12} b^6 ( -
b (
1440 (684k + 462 c - 1727 d) p + 385641132 k q + 4095636 c q 
\cr
-
12983877 d q + 94219368 k t + 9469882 c t - 66033204 d t) 
\cr
-
15 ( - 96 (520k + 43 c - 104 d) p -
 34077 c q + 136692 d q + 1857640 k t 
\cr
- 22998 c t +
47306 d t)) 
\cr
+
a^{15} b^3 (
-96 (-9324k + 2 c + 387 d) p + 22208160 kq + 500691 c q 
\cr
-
1942140 d q + 239709600 kt + 5842382 c t - 16780882 d t) +
15 ( 96 (-20k + d) p 
\cr
-
1116 c q + 3735 d q - 809096 k t - 14066 c t + 63272 d t) 
\cr
+
a^8 b^{10} (
96 (-1520k + 511 c - 134 d) p - 229047 c q 
\cr
-
63054 d q + 661742 c t - 567764 d t) +
b ( -
192 (-63332k + 8725 c + 7387 d) p 
\cr
+
6143760 c q + 14083365 d q + 160389568 k t - 33349956 c t +
1211602 d t) 
\cr
+
a^{14} b^4 ( - 96 (180k + 2 c - 9 d) p -
620448 k q - 14901 c q + 52512 d q + 910496 k t 
\cr
+ 17798 c t -
131728 d t) +
b (
96 (38700k + 781 c - 2491 d) p + 213241020 kq  + 4928781 c q 
\cr
-
16835793 d q - 178959408 k t + 2472926 c t + 11129654 d t) 
\cr
-
a^7 b^{11} (
15 (
96 (1220k + 383 c - 439 d) p - 2448324 kq 
\cr
- 261879 c q +
501360 d q + 4165984 kt + 202016 c t - 700056 d t) 
\cr
+b (
480 (-1616k+ 4568 c - 1991 d) p + 38861664 kq + 13797912 c q
\cr
 -
13942029 d q - 86969480 kt - 23944502 c t + 25663916 d t)) 
\cr
+
a^9 b^9 ( -
672 (-400k + 29 c + 62 d) p - 3995256 k q - 199923 c q 
\cr
+
725346 d q + 8950184 k t - 237858 c t - 907442 d t) 
\cr
+
b ( -
96 (119620k + 9213 c - 40363 d) p + 49804440 k q + 2624862 c q 
\cr
-
20884221 d q - 231609256 k t + 22319212 c t + 37755238 d t) 
\cr
-
a^{11} b^7 ( 480 (424k + c - 8 d) p -
2963400 k q - 187257 c q 
}$$
\vfill\eject
\eqn\grav{\eqalign{
+ 419913 d q + 9062904 k t +
270722 c t - 497360 d t) 
\cr
+
b (
384 (-60401k + 1196 c + 6831 d) p 
\cr
+ 461626416 kq +
17580171 c q - 64284969 d q 
\cr
- 722013808 k t + 14928440 c t +
42684832 d t) 
\cr
-
2 a^{17} b^2 (
(-3097048k - 244028 c + 464157 d) t 
\cr
-
2 a^5 b^{13} ( 96 (4k + 9 c - 5 d) p -
58884 kq - 8973 c q 
\cr
+ 11847 d q + 70960 kt + 11540 c t -
14640 d t) +
65 b ( -
2 (144 (20k + 3 c - 4 d) p 
\cr
+
9 (3604k + 743 c - 822 d) q - 32588 kt - 6995 c t +
7571 d t))) 
\cr
+
a^{13} b^5 (( - 96 (-520k + 11 c - 4 d) p -
790128 kq - 44247 c q 
\cr
+ 73722 d q + 4384696 kt + 124148 c t -
207996 d t) +
\cr
b (
2 (55580k + 48 (1865 c - 3 (43676k + 379 d)) p 
\cr
+
6 (16303472k + 1230680 c - 2118925 d) q 
\cr
- 407402424 kt -
14101721 c t + 23167123 d t))) 
\cr
+
a^6 b^{12} ( -
15 ( - 480 (204k + 5 c - 33 d) p +
633804 k q - 115959 c q + 4059 d q 
\cr
- 338248 k t + 265498 c t -
122678 d t) 
\cr
+
b ( -
2 (96 (15676k + 5219 c - 4699 d) p -
3 (5168668k + 279987 c - 754942 d) q 
\cr
+
20 ((913672k + 50071 c - 134207 d) t))))
\rbrack 
\rbrace+{\lbrace}a\leftrightarrow{b}{\rbrace}}}
Next, we perform  the conformal transformation
from the half-plane coordinates $z\equiv{a},{\bar{z}}\equiv{b}$
to the $r,\alpha$ coordinates in the disc
($0\leq{r}\leq{1},0\leq\alpha\leq{2\pi}$) using the prescription
(58)-(59)
and replacing each of the half-plane integrals of the type
$\int{da}{db}(a-b)^{\gamma_1}a^{\gamma_2}b^{\gamma_3}$
with corresponding integrals on the disc. 
Straightforward evaluation
of the asymptotics of the disc integrals in the field theory limit
$\alpha^\prime\rightarrow{0}$ then gives leads to (60) , skipping
the contact terms.
Finally, the lower-derivative contributions to the 
correlator (46) are given by the overall expession (61) that 
vanishes in four dimensions.
For completeness, we finish with presenting explicit
expressions for $I_{lower-der}(k_1,k_2,k_3)$ entering (61).
It is convenient to cast it according to 
\eqn\grav{\eqalign{
I_{lower-der}(k_1,k_2,k_3)
=I_3+I_5+I_7}}
with $I_p$ being the $p$-derivative pieces.
The integration procedure is then identical to the one explained
for the 9-derivative contribution.
Introducing further convenient abbreviations:
\eqn\grav{\eqalign{
u_1\equiv\omega_{m_1}^{a_1b_1}(k_2)\omega_{m_2}^{a_2b_2}(k_3)
\omega_{m_3}^{a_3b_3|cd}(k_1)\eta^{a_1a_2}(k_1^a+k_2^a+k_3^a)
(k_1^{b_2}+k_2^{b_2}+k_3^{b_2})k_1^{b_1}
\cr
u_2\equiv\omega_{m_1}^{a_1b_1}(k_2)\omega_{m_2}^{a_2b_2}(k_3)
\omega_{m_3}^{a_3b_3|cd}(k_1)\eta^{a_1a_2}(k_1^a+k_2^a+k_3^a)
(k_1^{b_2}+k_2^{b_2}+k_3^{b_2})k_3^{b_1}
\cr
v_1\equiv\omega_{m_1}^{a_1b_1}(k_2)\omega_{m_2}^{a_2b_2}(k_3)
\omega_{m_3}^{a_3b_3|cd}(k_1)\eta^{a_1a}
\cr\times
(k_1^{a_2}k_1^{b_2}+k_2^{a_2}k_1^{b_2}
+k_1^{a_2}k_2^{b_2}+k_2^{b_2}k_2^{a_2})
k_1^{b_1}\cr
v_2\equiv\omega_{m_1}^{a_1b_1}(k_2)\omega_{m_2}^{a_2b_2}(k_3)
\omega_{m_3}^{a_3b_3|cd}(k_1)\eta^{a_1a}
\cr\times
(k_1^{a_2}k_1^{b_2}+k_2^{a_2}k_1^{b_2}
+k_1^{a_2}k_2^{b_2}+k_2^{b_2}k_2^{a_2})
k_2^{b_1}\cr
v_3\equiv\omega_{m_1}^{a_1b_1}(k_2)\omega_{m_2}^{a_2b_2}(k_3)
\omega_{m_3}^{a_3b_3|cd}(k_1)\eta^{a_1a}
\cr\times
(k_1^{a_2}k_1^{b_2}+k_2^{a_2}k_1^{b_2}
+k_1^{a_2}k_2^{b_2}+k_2^{b_2}k_2^{a_2})
k_3^{b_1}\cr
\lambda_1\equiv\omega_{m_1}^{a_1b_1}(k_2)\omega_{m_2}^{a_2b_2}(k_3)
\omega_{m_3}^{a_3b_3|cd}(k_1)\eta^{a_1a_3}
\cr\times(k_1^a+k_2^a+k_3^a)
(k_1^{a_2}k_1^{b_2}+k_2^{a_2}k_1^{b_2}
+k_1^{a_2}k_2^{b_2}+k_2^{b_2}k_2^{a_2})
k_1^{b_1}k_2^{b_3}\cr
\lambda_2\equiv\omega_{m_1}^{a_1b_1}(k_2)\omega_{m_2}^{a_2b_2}(k_3)
\omega_{m_3}^{a_3b_3|cd}(k_1)\eta^{a_1a_3}
\cr\times(k_1^a+k_2^a+k_3^a)
(k_1^{a_2}k_1^{b_2}+k_2^{a_2}k_1^{b_2}
+k_1^{a_2}k_2^{b_2}+k_2^{b_2}k_2^{a_2})
k_3^{b_1}k_2^{b_3}\cr
\lambda_3\equiv\omega_{m_1}^{a_1b_1}(k_2)\omega_{m_2}^{a_2b_2}(k_3)
\omega_{m_3}^{a_3b_3|cd}(k_1)\eta^{a_1a_3}
\cr\times(k_1^a+k_2^a+k_3^a)
(k_1^{a_2}k_1^{b_2}+k_2^{a_2}k_1^{b_2}
+k_1^{a_2}k_2^{b_2}+k_2^{b_2}k_2^{a_2})
k_1^{b_1}k_3^{b_3}\cr
\lambda_4\equiv\omega_{m_1}^{a_1b_1}(k_2)\omega_{m_2}^{a_2b_2}(k_3)
\omega_{m_3}^{a_3b_3|cd}(k_1)\eta^{a_1a_3}
\cr\times(k_1^a+k_2^a+k_3^a)
(k_1^{a_2}k_1^{b_2}+k_2^{a_2}k_1^{b_2}
+k_1^{a_2}k_2^{b_2}+k_2^{b_2}k_2^{a_2})
k_3^{b_1}k_3^{b_3}\cr
\rho_1\equiv\omega_{m_1}^{a_1b_1}(k_2)\omega_{m_2}^{a_2b_2}(k_3)
\omega_{m_3}^{a_3b_3|cd}(k_1)\eta^{aa_3}
\cr\times
(k_1^{a_2}k_1^{b_2}+k_2^{a_2}k_1^{b_2}
+k_1^{a_2}k_2^{b_2}+k_2^{b_2}k_2^{a_2})k_2^{b_3}
\cr
\rho_2\equiv\omega_{m_1}^{a_1b_1}(k_2)\omega_{m_2}^{a_2b_2}(k_3)
\omega_{m_3}^{a_3b_3|cd}(k_1)\eta^{aa_3}
\cr\times
(k_1^{a_2}k_1^{b_2}+k_2^{a_2}k_1^{b_2}
+k_1^{a_2}k_2^{b_2}+k_2^{b_2}k_2^{a_2})k_3^{b_3}
}}
the asymptotics of the 7-derivative contribution is computed to give
\vfill \eject
\eqn\grav{\eqalign{I_7
\sim
(k_1k_2)^{-1}
(k_1k_3)^{-1}
(k_2k_3)^{-1}
\cr\times
\lbrace
u_1(1493472p+1195776q+505040t)+g(2051748p+
2154326q+908764t)
\cr
+h(2214564p+4140804q+764500t)
-u_2(f(3649254p+1628060q+492190t)
\cr
+
g(3166474p+1040800q+30605620t)+h(2082610p-2427639q+
1648244t))
\cr
+v_1(f(132128p+2003462q+808692t)+g(988622p+8267294q+
3016884t)
\cr
+h(-4428732p+10608920q-764824t))
+v_2(f(8467662p+5883454q+936720t)
\cr
+g(12845628p+
9832562q+964355t)+h(20605796p+1557672q-2062380t))
\cr
+v_3(f(7884240p-3167390q-1010042t)+g(8690112p+
40306q+2087600t)
\cr
+h(-3860368p-894766q+7057t))
-\lambda_1(9085p-30492495q+735944t)
\cr
+
\lambda_2( 2235256p-41879080q-538906t)
+\lambda_3(1264326p+24246476q+426t)+
\cr
\lambda_4(130p+284472q+3280845t)
+\rho_1(c(20047925p+2823510q+724612t)
\cr
+d(2240244p+3171313q+462884t)+k(1021245p+2436124q+6523480t))
\cr
+\rho_2(c(-4086001p+2049050q+206800t)+d(2006114p
-929683q+96492t)
\cr
+k(-24p+907200q+1920160t))
\rbrace+...}}
Next, to describe the $5$-derivative piece, we shall adopt
the notations:
\eqn\grav{\eqalign{
\gamma_1\equiv\omega_{m_1}^{a_1b_1}(k_2)\omega_{m_2}^{a_2b_2}(k_3)
\omega_{m_3}^{a_3b_3|cd}(k_1)\eta^{a_1a_2}\eta^{aa_3}(k_1^{b_2}+k_2^{b_2})
k_1^{b_1}k_2^{b_3}
\cr
\gamma_2\equiv\omega_{m_1}^{a_1b_1}(k_2)\omega_{m_2}^{a_2b_2}(k_3)
\omega_{m_3}^{a_3b_3|cd}(k_1)\eta^{a_1a_2}\eta^{aa_3}(k_1^{b_2}+k_2^{b_2})
k_3^{b_1}k_2^{b_3}
\cr
\gamma_3\equiv\omega_{m_1}^{a_1b_1}(k_2)\omega_{m_2}^{a_2b_2}(k_3)
\omega_{m_3}^{a_3b_3|cd}(k_1)\eta^{a_1a_2}\eta^{aa_3}(k_1^{b_2}+k_2^{b_2})
k_1^{b_1}k_3^{b_3}
\cr
\gamma_4\equiv\omega_{m_1}^{a_1b_1}(k_2)\omega_{m_2}^{a_2b_2}(k_3)
\omega_{m_3}^{a_3b_3|cd}(k_1)\eta^{a_1a_2}\eta^{aa_3}(k_1^{b_2}+k_2^{b_2})
k_2^{b_1}k_3^{b_3}
\cr
\delta_1\equiv\omega_{m_1}^{a_1b_1}(k_2)\omega_{m_2}^{a_2b_2}(k_3)
\omega_{m_3}^{a_3b_3|cd}(k_1)\eta^{b_1b_3}\eta^{aa_1}
\cr\times
(k_1^{a_2}k_1^{b_2}+k_2^{a_2}k_1^{b_2}
+k_1^{a_2}k_2^{b_2}+k_1^{b_2}k_1^{a_2})k_2^{a_3}
\cr
\delta_2\equiv\omega_{m_1}^{a_1b_1}(k_2)\omega_{m_2}^{a_2b_2}(k_3)
\omega_{m_3}^{a_3b_3|cd}(k_1)\eta^{b_1b_3}\eta^{aa_1}
\cr\times
(k_1^{a_2}k_1^{b_2}+k_2^{a_2}k_1^{b_2}
+k_1^{a_2}k_2^{b_2}+k_1^{b_2}k_1^{a_2})k_3^{a_3}
\cr
\epsilon_1\equiv\omega_{m_1}^{a_1b_1}(k_2)\omega_{m_2}^{a_2b_2}(k_3)
\omega_{m_3}^{a_3b_3|cd}(k_1)\eta^{b_1b_3}\eta^{a_1a_3}
\cr\times
(k_1^{a_2}k_1^{b_2}+k_2^{a_2}k_1^{b_2}
+k_1^{a_2}k_2^{b_2}+k_1^{b_2}k_1^{a_2})(k_1^a+k_2^a+k_3^a)
}}
Then
the asymptotics of the 5-derivative contribution is computed to give
\eqn\grav{\eqalign{I_5
\sim
(k_1k_2)^{-1}
(k_1k_3)^{-1}
(k_2k_3)^{-1}
\cr\times
\lbrace
\gamma_1(1864828p+
2866523q-303t)+\gamma_2(20078254p+652q-
1887348t)
\cr
+\gamma_3(1542620p-1288760q-1052t)+
\gamma_4(30884240p+6268906q+9208t)
\cr
+
\delta_1(2305266p+764080q-2006875t)+
\delta_2(2640884p+306708q+9861808t)
\cr
+
\epsilon_1(3077454p-708616q-979072t)
\rbrace+...}}
Finally, to describe the $3$-derivative contributions
(up to contact terms)
we denote
\eqn\grav{\eqalign{
\sigma_1\equiv\omega_{m_1}^{a_1b_1}(k_2)\omega_{m_2}^{a_2b_2}(k_3)
\omega_{m_3}^{a_3b_3|cd}(k_1)\eta^{a_1a_2}\eta^{aa_3}\eta^{b_1b_2}
k_2
\cr
\sigma_2\equiv\omega_{m_1}^{a_1b_1}(k_2)\omega_{m_2}^{a_2b_2}(k_3)
\omega_{m_3}^{a_3b_3|cd}(k_1)\eta^{a_1a_2}\eta^{aa_3}\eta^{b_1b_2}
k_3}}
Then
the asymptotics of the 3-derivative contribution is computed to give
\eqn\grav{\eqalign{I_3\sim
(k_1k_2)^{-1}
(k_1k_3)^{-1}
(k_2k_3)^{-1}
\cr\times
\lbrace
\sigma_1(-10887660p+30458284q+41010562t)
\cr
+
\sigma_2(10865492p-30460944q-41010934t)
\rbrace+...
}}
This concludes the evaluation of $I(k_1,k_2,k_3)$
factor describing the lower derivative contributions to
the correlator (46), modulo contact terms.

\listrefs

\end